\documentclass[journal,transmag]{IEEEtran}

\usepackage{array}
\usepackage{bm}
\usepackage{bbm}
\usepackage{bbold}
\usepackage{graphicx}
\usepackage{subcaption}
\usepackage{url}
\usepackage{placeins}
\usepackage{algpseudocode}
\usepackage{amsmath}
\usepackage{amssymb}
\usepackage{amsfonts}
\usepackage{amsthm}
\usepackage{color}
\usepackage{cite}
\usepackage{hyperref}
\usepackage[dvipsnames]{xcolor}

\newtheorem{lemma}{Lemma}
\newtheorem{theorem}{Theorem}
\newtheorem{corollary}{Corollary}

\usepackage{calc}
\usepackage{tikz}
\usetikzlibrary{decorations.markings, arrows}
\tikzstyle{vertex}=[circle, draw, inner sep=0pt, minimum size=6pt]

\newcommand{\mat}[1]{\mathbf{#1}}
\newcommand{\matt}[1]{\mathbf{#1}(t)}
\newcommand{\gvec}[1]{\bm{#1}}
\newcommand{\vecb}[1]{\mathbf{#1}}
\newcommand{\norm}[1]{\left\| #1 \right\|}

\newcommand{\tdt}{t+\Delta t}
\newcommand{\dP}{\Delta P}
\newcommand{\vpi}{\gvec{\pi}}

\newcolumntype{L}{>{\centering\arraybackslash}m{0.4\textwidth}}

\title{Tie-decay networks in continuous time and eigenvector-based centralities}

\author{\IEEEauthorblockN{Walid Ahmad\IEEEauthorrefmark{1},
Mason A. Porter\IEEEauthorrefmark{2}, and
Mariano Beguerisse-D\'iaz\IEEEauthorrefmark{1,3}}
\IEEEauthorblockA{\IEEEauthorrefmark{1}Mathematical Institute, University of Oxford, Oxford, UK}
\IEEEauthorblockA{\IEEEauthorrefmark{2}Department of Mathematics, University of California, Los Angeles, USA}
\IEEEauthorblockA{\IEEEauthorrefmark{3}Spotify Ltd, UK}
\thanks{Corresponding author: M. A. Porter (e-mail: \tt{mason@math.ucla.edu})}}

\begin{document}

\IEEEtitleabstractindextext{
\begin{abstract}
Network theory is a useful framework for studying interconnected
systems of interacting entities. Many networked systems evolve
continuously in time, but most existing methods for the analysis of
time-dependent networks rely on discrete or discretized time. In this
paper, we propose an approach for studying networks that evolve in
continuous time by distinguishing between \emph{interactions}, which
we model as discrete contacts, and \emph{ties}, which encode the
strengths of relationships as functions of time. To illustrate our
tie-decay network formalism, we adapt the well-known PageRank
centrality score to our tie-decay framework in a mathematically
tractable and computationally efficient way. We apply this framework to a synthetic example and
then use it to study a network of retweets during the 2012 National
Health Service controversy in the United
Kingdom. Our work also provides guidance for similar
generalizations of other tools from network theory to continuous-time
networks with tie decay, including for applications to streaming data.
 \end{abstract}
}

\maketitle


\IEEEdisplaynontitleabstractindextext

\IEEEpeerreviewmaketitle



\section{Introduction}
Networks provide a versatile framework to model and analyze complex
systems of interacting entities~\cite{newman2010networks}. In many
complex systems, interaction patterns change in time and the entities
can also leave or enter the system at different times. To accurately
model and understand such systems, it is essential to incorporate
temporal information about their interactions into network
representations~\cite{Katz1959, Farmer1987, Wasserman1988,
  Snijders2001, Carley2003, Belykh2004, braha2006centrality}. See Refs.~\cite{holme2012temporal,holme2015modern,masuda-book,holme2019} for
overviews of the study of time-dependent networks, which are often
also called \emph{temporal networks} or \emph{dynamic networks}.

A major challenge in the analysis of temporal networks is that one
often has to discretize time by aggregating connections into time
windows. Given a discrete or discretized set of interactions, one can
then analyze communities, important nodes, and other facets of such
networks by examining a multilayer-network representation of these
interactions~\cite{holme2015modern,kivela2014multilayer,Taylor2015,bazzi2016community}. 
An important challenge that arises with aggregation is that 
there may not be any obvious or even any `correct' size of a time window (even when such aggregation
employs non-uniform time
windows~\cite{Sulo2010,caceres2011,psorakis2012,psorakis2013probabilistic}).
A window that is too small risks missing important network structures 
(e.g., by construing a signal as noise), but using an overly
large window may obscure important temporal features.
(See \cite{fenn2012} for one discussion.)  Moreover, in many
social systems, interactions are
bursty~\cite{Beguerisse2010,holme2012temporal,kivela2015}, which is a
crucial consideration when aggregating
interactions~\cite{hoffmann2012generalized} and is potentially a major
cause of concern when using homogeneous time
windows~\cite{psorakis2012}. Bursty interactions not only present a
challenge when choosing the width of the time windows, but they also
challenge {where} to place the boundaries such windows. Shifting time windows
forward or backward may significantly alter the statistics of
a data set, even when one does not change the width of
the windows~\cite{kivela2015}.

From a modeling perspective, aggregating interactions often may not
be an appropriate approach for systems with asynchronous activity or
which evolve continuously in time. See~\cite{Valdano2018} for an
investigation of biological contagions, \cite{yang2018} for a study of
influential users in social networks, \cite{zino2016,zino2017} for a
generalization of the formalism of `activity-driven networks' to
continuous time, \cite{Motegi2012} for a study of rankings in
competitive sports, and~\cite{Flores2018} for a general
continuous-time framework for temporal networks. In many cases,
contacts in a temporal network can have a noninstantaneous duration, and it can be important
to take such information into account~\cite{Onnela2007,moro2015}. For
example, the phone-call data that were studied in~\cite{Grindrod2014}
require contacts to exist for the duration of a phone call. In other
cases, interactions can be instantaneous (e.g., a mention in a tweet,
a text message, and so on), and their importance decreases over
time~\cite{Burt2000,laub2015}. For many types of temporal networks
(e.g., feeds on social media), there is also a decay in attention span
for reacting to
posts~\cite{Hodas2012,lerman-digg2,lerman2016information}.

In the present paper, we introduce a framework for modeling temporal
networks in which the strength of a connection (i.e., a tie) can
evolve continuously in time. For example, perhaps the strength of a
tie decays exponentially after the most recent interaction. (One can also use point-process models like Hawkes processes \cite{laub2015} to examine similar ideas from a node-centric perspective.) Our mathematical formalism of such `tie-decay networks' allows us to
examine them using analytical calculations and to implement them
efficiently in real-world applications with streaming data. We
showcase our tie-decay formalism by computing continuous-time PageRank
centrality scores for both a synthetic temporal network and a temporal
network that we construct from a large collection of Twitter
interactions over the course of a year.

Our paper proceeds as follows. In Section~\ref{ties},
we formalize our discussion of ties, interactions, and temporal
networks. We also introduce the notion of \emph{tie-decay networks},
which is the focus of our study. In Section~\ref{evec}, we formulate
how to study eigenvector-based centralities in tie-decay networks.
In Section~\ref{sec:synthetic}, we construct a synthetic network with
known properties to illustrate some of the pitfalls of binning
interactions and how tie-decay networks can avoid them. In
Section~\ref{nhs}, we discuss and compute tie-decay PageRank
centralities to examine important agents in a National Health Service
(NHS) retweet network. In Section~\ref{conclude}, we conclude and
discuss the implications of our work. We give proofs of our main
theoretical results in Appendix~\ref{sec:proof_boundPR}.


\section{Ties, Interactions, and Temporal Networks} \label{ties}
We seek to construct a continuous-time temporal network that can
capture the evolution of relationships between entities in a network. To do this, we
make an important distinction between `interactions' and `ties'. An
\emph{interaction} between two entities is an event that takes place at
a specific time interval or point in time (e.g., a face-to-face
meeting, a text message, or a phone call). By contrast, a \emph{tie}
between two entities is a relationship between them. A tie between two entities can have a
weight to represent its strength (such as the strength of a friendship
or collaboration). Ties between entities strengthen with repeated
interactions, but they can also deteriorate in their
absence~\cite{Burt2000,moro2017,michalski2018}. There are many
empirically plausible, domain-specific deterioration (i.e.,``decay") functions
 that one can use; examples include linear decay, power-law decay, 
and exponential decay~\cite{Burt2000,moro2017,michalski2018,wixted1991form}.
In the present paper, we use exponential decay, which is a common choice for the intensity decay function in Hawkes processes~\cite{laub2015}.
We restrict ourselves to modeling instantaneous interactions, but it
is possible to generalize our tie-decay formalism to incorporate interactions with different durations.

Consider a set of $n$ interacting entities (i.e., nodes), and let $B(t)$ be the $n
\times n$ time-dependent, real, non-negative matrix whose entries
$b_{ij}(t)$ represent the connection strengths between entities $i$ and $j$
at time $t$.  To construct a continuous-time temporal network of these ties,
we make two modeling assumptions about how ties evolve and how
interactions strengthen them:
\begin{enumerate}
\item[(1)]{In the absence of interactions, we assume that ties decay
  exponentially, as proposed by {Jin et
    al.}~\cite{Jin2001}. In mathematical terms, $b_{ij}'=-\alpha
  b_{ij}$ (where the prime represents differentiation with respect to
  time), so $b_{ij}(t)=b_{ij}(0)e^{-\alpha t}$ for some $\alpha \geq
  0$ and an initial condition $b_{ij}(0)$.}
\item[(2)]{If two entities interact at time $t=\tau$, the strength of the tie
  between them grows instantaneously by $1$, and it then decays as normal. This
  choice differs from~\cite{Jin2001}, who reset the strength to $1$
  after each interaction.}
\end{enumerate}
Taken together, these assumptions imply that the temporal evolution
of a tie satisfies the ordinary differential equation (ODE)
\begin{equation}
	  b_{ij}' = -\alpha b_{ij} + \delta(t - \tau)e^{-\alpha(t - \tau)}\,.
  \label{eq:tie_ode}
\end{equation}
In equation \eqref{eq:tie_ode}, we represent an instantaneous
interaction at $t=\tau$ as a pulse with the Dirac
$\delta$-function. If the tie has the resting initial condition
$b_{ij}(0)=0$, the solution to equation~\eqref{eq:tie_ode} is
$b_{ij}(t) = H(t-\tau)e^{-\alpha(t-\tau)}$, where $H(t)$ is the
Heaviside step function. This formulation is related to the one in
Flores and Romance~\cite{Flores2018}, who integrated over functions
that represent the temporal evolution of interactions. A related notion of tie decay appears in the
  work of Sharan and Neville~\cite{Sharan2007}, although
  they applied decay to a sequence of graphs in discrete time,
  instead of individual edges in continuous time. When there are multiple
interactions between entities, we represent them as streams of pulses in
the $n\times n$ matrix $\widetilde{A}(t)$. If entity $i$ interacts with
entity $j$ at times $\tau_{ij}^{\left(1\right)},\,
\tau_{ij}^{\left(2\right)}, \,\ldots$, then $\tilde{a}_{ij}(t) =
\sum_{k}\delta(t-\tau_{ij}^{\left(k\right)})e^{-\alpha(t -
  \tau_{ij}^{\left(k\right)})}$.  We rewrite
equation~\eqref{eq:tie_ode} as
\begin{equation}
  	b_{ij}' = -\alpha b_{ij} + \tilde{a}_{ij}\,,
  \label{eq:tie_ode2}
\end{equation}
 which has the solution $b_{ij}(t) = \sum_k
 H(t-\tau_{ij}^{\left(k\right)})e^{-\alpha
   (t-\tau_{ij}^{\left(k\right)})}$ from a resting\footnote{It is not
   unlike a Norwegian blue parrot. (Norwegian blues stun easily.)}
 initial condition.

In practice --- and, specifically, in data-driven applications --- one
can readily construct $B(t)$ by discretizing time so that there is at
most one interaction during each time step of length $\Delta t$ (e.g.,
in a Poisson process). Such time discretization is common in the
simulation of stochastic dynamical systems, such as Gillespie
algorithms~\cite{Erban2007,porter2016dynamical,vester2015}. In our
case, we let ${A}(t)$ be the $n \times n$ matrix in which an entry
$a_{ij}(t)=1$ if entity $i$ interacts with entity $j$ at time $t$ and
$a_{ij}(t)=0$ otherwise. At each time step, $A(t)$ has at most one
nonzero entry for a directed network (and at most two of them for an
undirected network). Therefore,
\begin{equation}
	  {B}(\tdt)  = e^{-\alpha \Delta t}B(t) + A(\tdt)\,.
  \label{eq:Bupdate}
\end{equation}
Equivalently, if interactions between pairs of entities occur at times
$\tau^{\left(\ell\right)}$ (it can be a different pair at different
times) such that $0\leq \tau^{\left(0\right)} < \tau^{\left(1\right)}
< \ldots < \tau^{\left(T\right)}$, then at $t\geq
\tau^{\left(T\right)}$, we have
\begin{equation}
	  B(t) = \sum_{k = 0}^T e^{-\alpha (t - \tau^{\left(k\right)})} A(\tau^{\left(k\right)})\,.
  \label{eq:B_summation}
\end{equation}
If there are no interactions at time $t$, then every entry of the matrix $A(t)$ is $0$.

Our continuous-time approach avoids having to impose a hard partition
of the interactions into bins (i.e., windows). 
However, one still needs to choose a
value for the decay parameter $\alpha$.  Another benefit of our
approach is that it eliminates the placement of the time windows as a
potential source of bias \cite{kivela2015}. When choosing a value for
$\alpha$, it is perhaps intuitive to think about the {\it half-life}
$\eta_{1/2}$ of a tie, as it gives the amount of time for a tie to
lose half of its strength in the absence of new interactions. Given
$\alpha >0$, the half-life of a tie is $\eta_{1/2} =
\alpha^{-1}\ln{2}$. Our choice of using $\alpha$ to downweight old
activity is consistent with \cite{grindrod2013matrix, Grindrod2014},
which used a similar exponential decay factor to filter out older
interactions in the context of dynamic communicability.

\begin{figure}[tp]
  \centerline{\includegraphics[width=0.45\textwidth]{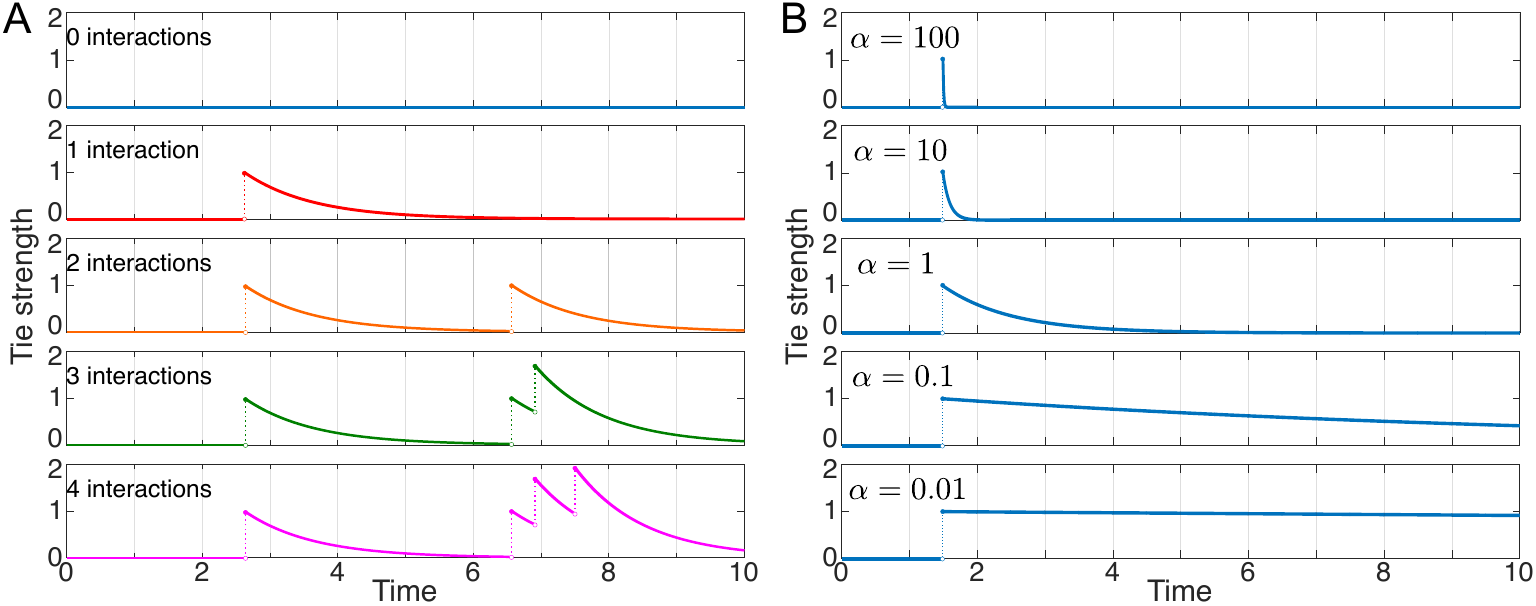}}
  \caption{{\bf A:} Evolution of strength of a tie with exponential decay [see equation~\eqref{eq:tie_ode2}]. {\bf B:} The decay rate
    $\alpha$ determines how fast the tie decays.
    }
  \label{fig:toy_example}
\end{figure}

In Fig.~\ref{fig:toy_example}A, we illustrate the evolution of the strength of a tie
in our tie-decay formalism.
If entities $i$ and $j$ have never
interacted before time $t_0$, then $b_{ij}(t_0) = 0$. Suppose that
they first interact at time $\tau^{\left(1\right)} > t_0$ (such that
$a_{ij}(\tau^{\left(1\right)}) = 1$). Their tie strength then
increases by $1$, so $b_{ij}(\tau^{\left(1\right)}) = 1$. It
subsequently decays exponentially until they interact again, so $b_{ij}(t>\tau^{\left(1\right)}) =
e^{-\alpha (t - \tau^{\left(1\right)})}$ before their next interaction. If entities $i$ and $j$ next
interact at time $\tau^{\left(2\right)} > \tau^{\left(1\right)}$, such
that $a_{ij}(\tau^{\left(2\right)})=1$, their tie strength becomes
$b_{ij}(\tau^{\left(2\right)}) = e^{-\alpha (\tau^{\left(2\right)} -
  \tau^{\left(1\right)})} + 1$, and so on.


\section{Eigenvector-Based Centrality Scores in Tie-Decay Networks}
\label{evec}
  One common question that
arises frequently when analyzing a network in
  scientific and industrial applications is the following: {\it What
  are the most important nodes?}~\cite{Grindrod2016} To examine this
question, researchers have developed numerous {\it centrality scores}
to quantify the importance of nodes according to different
criteria~\cite{newman2010networks}.

An important family of centrality scores arise from spectral
properties of the adjacency matrix (or other matrices) of a
network~\cite{Masuda2016,Bonacich2001,Perra2008,Taylor2015}. One
attractive feature of computing centrality scores using a spectral method is that one can exploit the full structure of a network. 
By contrast, degree centrality is a simple centrality score that relies only on a network's local structure.
Eigenvector-based centrality scores have been insightful in numerous
applications, and one can use efficient numerical algorithms to
compute eigenvectors and singular vectors of
matrices~\cite{Golub1996,Trefethen1997}. Some of the most widely-used
spectral centrality scores for directed networks include
PageRank~\cite{Page1998,Gleich2015} (which exploits the properties of
a random walk on a network) and hub and authority
scores~\cite{Kleinberg1999} (which exploit both random-walk properties
and the asymmetry of connections in directed networks).

In temporal networks, centrality scores must incorporate not only
which nodes and edges are present in a network, but also when they are
present~\cite{Kim2012, Pan2011}. This makes it challenging to develop and analyze
centrality measures in temporal networks. Some approaches have
exploited numerical methods for dynamical systems to compute specific
scores, such as a Katz centrality for temporal
networks~\cite{Grindrod2014, Beres2018}, and others have employed
aggregated or multilayer representations of temporal networks to
calculate spectral centrality scores~\cite{Beguerisse2017,
  Taylor2015}. However, these approaches have either been limited to a
specific kind of centrality, or they have relied on the judicious
aggregation of interactions into time bins.
For example, an early paper~\cite{braha2006centrality} on centralities in temporal networks used a time bin of one day.
 Choosing an appropriate size for such a time bin is far from straightforward and requires deep
knowledge of the system under study: overly coarse bins
obfuscate temporal features, whereas bins that are too small may
obscure network structures, yielding scores that may result more
from noise than from signals.

Our tie-decay network formalism in equation~\eqref{eq:tie_ode2}
allows us to employ efficient numerical techniques to compute a
variety of spectral centrality scores in our temporal networks.  One
can tune the decay parameter $\alpha$ (which one can also generalize
to be node-specific, tie-specific, or time-dependent) to consider
different time scales.  A key benefit of our approach is that we
can easily incorporate both new interactions and new nodes as a
network evolves. In the present paper, we showcase an application
using PageRank centrality, but it is also worthwhile to study other spectral centrality scores using our tie-decay formalism.


\subsection{Tie-Decay PageRank Centrality}

PageRank centrality is a widely used (and historically important)
eigenvector-based centrality score for time-independent
networks~\cite{Page1998,newman2010networks}. The PageRank score of a
node in a network corresponds to its stationary distribution in a
random walk with teleportation~\cite{Gleich2015, Masuda2016}. In this type of random
walk, a walker continues its walk from a node by following an outgoing edge with
probability $\lambda\in (0,1)$ (where, in most versions, one chooses
the edge with a probability that is proportional to its weight), and
it `teleports' to some other node in the network with probability
$1-\lambda$. It is common to choose the destination node uniformly at
random, but many other choices are
possible~\cite{Lambiotte2012,Gleich2015}. In the present paper, we
employ uniform teleportation with $\lambda=0.85$ (which is a
common choice). Let $B$ be the adjacency matrix of a weighted network
with $n$ nodes, so $b_{ij}$ encodes the weight of a directed tie from
node $i$ to node $j$.  The $n\times 1$ vector $\vpi$ of PageRank
scores, with $\vpi > 0$ and $\norm{\vpi}_1 = 1$, is the
leading-eigenvector solution of the eigenvalue problem
\begin{equation}
	  G^T\vpi = \vpi\,,
  \label{eq:evecG}
\end{equation}
where $G$ is the $n \times n$ rate matrix of a teleporting random
walk:
\begin{align}
 	 G & = \lambda \left(D^\dagger B  + \vecb{c}\vecb{v}^T\right)
  + (1-\lambda)\mathbb{1}\vecb{v}^T \label{eq:G} \\
  & = \lambda P + (1-\lambda) \mathbb{1}\vecb{v}^T\,, \notag
\end{align}
where $P= D^\dagger B + \vecb{c}\vecb{v}^T$; the matrix $D$ is the
diagonal matrix of weighted out-degrees, so $d_{ii}=\sum_k b_{ik}$ and
$d_{ij}=0$ when $i\neq j$; and $D^\dagger$ is its Moore--Penrose
pseudo-inverse.  The $n \times 1$ vector $\vecb{c}$ is an indicator of
`dangling nodes' (i.e., nodes with $0$ out-degree): $c_i =1-
d^\dagger_{ii}\sum_k b_{ik}$, so $c_i=1$ if the out-degree of $i$ is
$0$ and $c_i=0$ otherwise. Additionally, $\mathbb{1}$ is the $n\times
1$ vector of $1$s, and the $n \times 1$ distribution vector $\vecb{v}$
encodes the probabilities of each node to receive a teleported
walker. In the present paper, we use $v_i={1}/{n}$ for all $i$.

The perturbations to $D^\dagger B$ from $\vecb{v}$ and $\vecb{c}$
ensure the ergodicity of the teleporting random walk,
so the  Perron--Frobenius theorem guarantees that
  $G^T$ has a unique right leading eigenvector
$\vpi$ whose entries are all
strictly positive. To calculate
$\vpi$, one can perform a power iteration on
$G^T$~\cite{Trefethen1997}, but in practice we do not need to
explicitly construct $G^T$. The iteration
\begin{equation}
	  \vpi^{\left(k+1\right)} = \lambda P^T\vpi^{\left(k\right)} + (1-\lambda)\vecb{v}\,,
  \label{eq:iter_pr}
\end{equation}
with $\vpi^{\left(0\right)}=\vecb{\left(0\right)}$ or
$\vpi^{\left(0\right)}=\vecb{v}$, converges to $\vpi$ and preserves
the sparsity of $P$. This choice, which ensures that computations are
efficient, is equivalent to a power iteration~\cite{Gleich2015}.

To compute time-dependent PageRank scores from the tie-strength matrix
$B(t)$, we define the temporal transition matrix
\begin{equation}
 	 P(t) = D^{\dagger}(t)B(t) + \vecb{c}(t)\vecb{v}^T\,,
\label{eq:Ptemp}
\end{equation}
where $D(t)$ is the diagonal matrix of weighted out-degrees (i.e., the
row sums of $B(t)$) at time $t$. The rank-$1$ correction
$\vecb{c}(t)\vecb{v}^T$ depends on time, because the set of dangling
nodes can change in time (though $\vecb{v}$ remains
fixed)\footnote{Strictly speaking, once a node leaves a dangling-node set,
    it never returns. Although the tie strength decays exponentially,
    it never quite reaches $0$. In practice, one can
      opt to remove ties with $b_{ij}\ll 1$ to maintain the sparsity
    of $B(t)$. When one does this, nodes can return to the dangling-node
    set.}. The iteration to obtain the
time-dependent vector of PageRank scores $\vpi(t)$ is now given by
\begin{equation}
 	 \vpi^{\left(k+1\right)}(t) = \lambda P^T(t) \vpi^{\left(k\right)}(t) 
  + (1-\lambda)\mat{v}\,.
  \label{eq:priteration}
\end{equation}

To understand the temporal evolution of $\vpi(t)$, we begin by
establishing some properties of the temporal transition matrix $P(t)$
in the following lemma.

\begin{lemma}
  \label{lemma:no_inter}
  When there are no new interactions between times $t$ and $\tdt$, the
  entries of $A(\tdt)$ are all $0$ and $P(\tdt) = P(t).$ If there is a
  single new interaction from node $i$ to node $j$, such that
  $a_{ij}(\tdt)=1$, then $P(\tdt) = P(t) + \dP$, where
  \begin{align}
    \dP &= \frac{1}{1+e^{-\alpha \Delta
        t}d_{ii}(t)}\vecb{e}_{i}\vecb{e}_{j}^T \notag \\
        	&\quad - \frac{1}
        {d_{ii}(t)\left(1+e^{-\alpha \Delta t}d_{ii}(t)\right)}
        \vecb{e}_{i}\vecb{e}_{i}^T B(t)
        -c_i(t)v_i\vecb{e}_i\mathbb{1}^T
  \label{eq:deltaP}
  \end{align}
  and $\vecb{e}_{i}$ and $\vecb{e}_{i}$, respectively, are the $i$-th and $j$-th
canonical vectors.
\end{lemma}
The first term in the right-hand side of equation~\eqref{eq:deltaP} is
a matrix whose only nonzero entry is the $(i,j)$-th term, the second
term is a rescaling of the $i$-th row of $B(t)$, and the
third term is the perturbation due to teleportation.  An important
implication of Lemma~\ref{lemma:no_inter} is that the PageRank scores do not change when there are no
new interactions, so $\vpi(\tdt)=\vpi(t)$. If each node or tie has different decay rates
(so that now we index $\alpha$ as $\alpha_i$ or $\alpha_{ij}$), then
this no longer has to be the case.

When there are new interactions, the following result sets an upper
bound on how much the PageRank scores can change.
\begin{theorem}
\label{thm:boundPR}
Suppose that there is a single interaction between times $t$ and
$\tdt$ from node $i$ to node $j$, such that the change $\dP$ in the
transition matrix satisfies equation~\eqref{eq:deltaP}. 
It follows that
\begin{align}
	  &\norm{\vpi(\tdt) - \vpi(t)}_1  \notag \\
	&\quad   \leq \frac{2\lambda}{1-\lambda} \min
  \left\lbrace \pi_{i}(t), \, \frac{1}{1+e^{-\alpha \Delta
      t}d_{ii}(t)} - \frac{c_i(t)}{2} \right\rbrace\,.
  \label{eq:combinedprmax}
\end{align}
\end{theorem}

We present two corollaries of Theorem~\ref{thm:boundPR}. 
\begin{corollary}
If $i$ is a dangling node at time $t$, then
\begin{equation}
	  \norm{\vpi(\tdt) - \vpi(t)}_1 
  \leq \frac{2\lambda}{1-\lambda} \min 
  \left\lbrace \pi_{i}(t), \frac{1}{2} \right\rbrace\,.
\end{equation}
\label{corr:dangling}
\end{corollary}

\begin{corollary}
If node $i$ has one or more outgoing edges at time $t$, then
\begin{align}
	  &\norm{\vpi(\tdt) - \vpi(t)}_1 \notag \\
  &\quad \leq \frac{2\lambda}{1-\lambda} \min 
  \left\lbrace \pi_{i}(t), \, 
    \frac{1}{1+e^{-\alpha \Delta t}d_{ii}(t)} \right\rbrace\,.
\end{align}
\label{corr:notdangling}
\end{corollary}

We give proofs of Lemma \ref{lemma:no_inter},
Theorem~\ref{thm:boundPR}, and Corollaries \ref{corr:dangling} and
\ref{corr:notdangling} in Appendix~\ref{sec:proof_boundPR}.


\subsubsection{Temporal Iteration}

To calculate the PageRank scores at time $\tdt$, we use the iteration
in equation~\eqref{eq:priteration} to update the PageRank vector using
$\vpi(t)$ as the initial value. That is,
\begin{equation} 
	\vpi^{\left(0\right)}(\tdt) =  \vpi(t)\,.
\end{equation}
The relative error of the computed PageRank vector at iteration $k$ is
\begin{equation} \label{def:rel_error_k}
  \norm{e_{\mathrm{rel}}^{\left(k\right)}}_1 = \norm{\vpi(\tdt) -
    \vpi^{\left(k\right)}(\tdt)}_1\,.
\end{equation} 
A result from \cite{Bianchini2005} (see their Theorem 6.1) implies that
$\norm{e_{\mathrm{rel}}^{\left(k\right)}}_1 \leq
\lambda^{k}\norm{e_{\mathrm{rel}}^{\left(0\right)}}_1\,$. 
The relation $\vpi^{\left(0\right)}(\tdt) = \vpi(t)$ and Theorem~\ref{thm:boundPR},
then imply that
\begin{align} \label{eq:rel_error_k}
          	\norm{e_{\mathrm{rel}}^{\left(k\right)}}_1\, & \leq \lambda^k
          \norm{\vpi(\tdt)-\vpi(t)}_1 \notag \\ 
          	&\leq \frac{2\lambda^{k+1}}{1-\lambda} \min \left\lbrace
          \pi_{i}(t), \, \frac{1}{1+e^{-\alpha \Delta t}d_{ii}(t)} -
          \frac{c_i(t)}{2} \right\rbrace\,. 
\end{align}
Therefore, we can select an error tolerance $\epsilon$ such that
$\norm{e_{\mathrm{rel}}^{\left(k^{*}\right)}}_1 \leq \epsilon$ for
some number $k^*$ of iterations. The value $k^*$ represents the
maximum number of iterations that we need to have
a relative error of at most $\epsilon$. We compute $k^*$ by calculating
{\tiny
\begin{equation}
  k^* = \frac{\ln(\epsilon) - \ln(2) + \ln(1-\lambda)
    - \ln\left(\min \left\lbrace \pi_{i}(t)\,, \,
    \frac{1}{1+e^{-\alpha \Delta t}d_{ii}(t)} - \frac{c_i(t)}{2}
    \right\rbrace \right)}{\ln(\lambda)} -1 \,.
\end{equation}}
In practice, we can instead track the residual after $k$
iterations~\cite{Gleich2015}:
\begin{align}
  \mat{r}^{\left(k\right)}(\tdt) &=
  \vpi^{\left(k+1\right)}(\tdt)-\vpi^{\left(k\right)}(\tdt) \notag
  \\ &= (1-\lambda)\vecb{v} - \left(I_n - \lambda
  P^T(\tdt)\right)\vpi^{\left(k\right)}(\tdt)\,.
\end{align}
We use this residual to bound the relative error, 
{\tiny
\begin{align}
  \norm{\vpi(\tdt) - \vpi^{\left(k\right)}(\tdt)}_1 &= \norm{
    \left(I_n - \lambda
    P^T(\tdt)\right)^{-1}\mat{r}^{\left(k\right)}(\tdt)}_1 \notag
  \\ &\leq
  \frac{1}{1-\lambda}\norm{\mat{r}^{\left(k\right)}(\tdt)}_1\,,
\end{align}}
and we thereby monitor the convergence of the iteration. In our
experiments, we always obtain $\|\vpi^{\left(k+1\right)}(\tdt) -
\vpi^{\left(k\right)}(\tdt)\|_{l_1} <10^{-6}$ in two iterations or
fewer (see Section~\ref{sec:efficiency}). See~\cite{chien2004} for an
alternative approach for approximating PageRank after adding a single edge.


\section{A Synthetic Example}
\label{sec:synthetic}

To illustrate some of the features of our tie-decay formalism,
we construct an example that illustrates the challenges of binning
interactions and how our approach can help overcome them.

\begin{figure}[tp]
  \centering
  \includegraphics[width=0.45\textwidth]{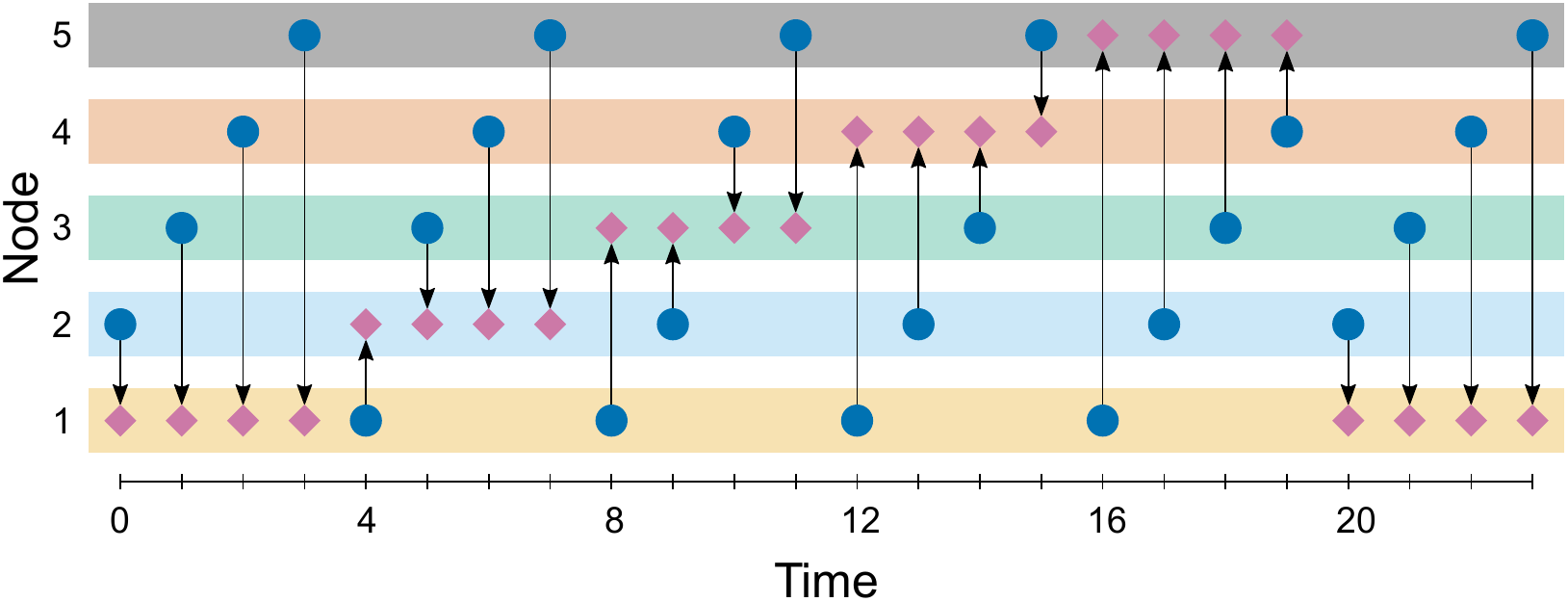}
  \caption{(Color online) Schematic illustration of the interactions in our
    synthetic network. (See the description in the main text.)
    We show target nodes as pink diamonds and
    source nodes as blue disks. In this example, there is one target at a
    time, and the different nodes take turns being the target. After
    all source nodes interact with the current target, a new node
    becomes the target, and all source nodes then interact with
    it. This cycle keeps repeating in this temporal network, which is
    periodic with a period of $20$ time steps.  
      }
  \label{fig:synNetTimeplot}
\end{figure}

Consider a network with node set $V = \lbrace{1, 2, 3, 4, 5}
\rbrace$. These nodes interact with each other cyclically in discrete
time, with only a single edge present during each time step.
 At the initial time $t_0$, we select a `target
node' and a `source node' and create a directed edge from the source
to the target. At the next time (i.e., $t = t_1$), there is an edge
from a different source node to the same target. This occurs with a
third source node and the same target at time $t_2$ and with the final
remaining source nodes and this target at time $t_3$. We have now
exhausted all possible source nodes for this target, so at time $t_4$,
we select a new target node and repeat the above
process. Specifically, there is exactly one directed edge at each discrete time,
and we select each of the four possible source nodes exactly
once between times $t_4$ and $t_7$. We repeat this cycle thrice more,
and we finish with the final source--target pair at time
$t_{19}$. The process begins again at time $t_{20}$ with the first target
node, and it continues periodically. In
Fig.~\ref{fig:synNetTimeplot}, we show the interactions in this
synthetic temporal network for the first $25$ time steps.  When we
examine temporal centralities in this network, we expect its cyclical
behavior to be reflected in the centrality scores of its nodes.

\begin{figure}[tp]
  \centering
  \includegraphics[width=.5\textwidth]{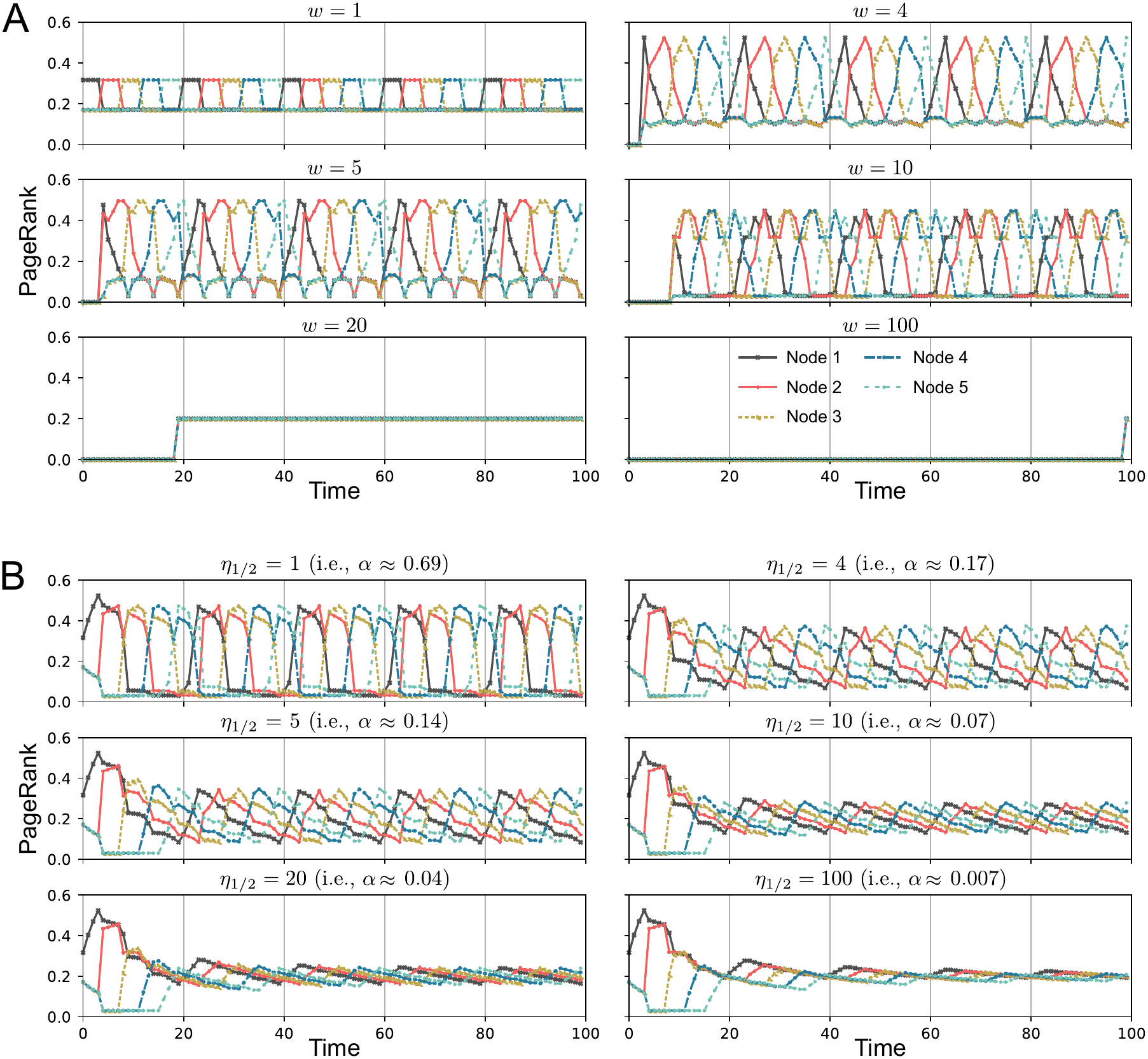}
  \caption{(Color online) 
  PageRank scores over time for each node in our five-node synthetic
    network. (See the description in the main text.)
     {\bf A:} Calculations of time-independent PageRank scores using sliding windows $(t_k - w, t_k]$ of
  different duration to aggregate interactions.  {\bf B:} Calculations
  of tie-decay PageRank scores for different values of the tie half-life.
  }
  \label{fig:synNetCent}
\end{figure}

One way to examine temporal centralities of the nodes in our synthetic
network is by aggregating the interactions into discrete sliding
windows of duration $w$~\cite{peel2015detecting, ridder2016detection}.
To do this, we construct time-independent networks with interactions in the time window $(t_k - w, t_k]$,
 and we then compute the PageRank
scores of the nodes for each of these networks. In
Fig.~\ref{fig:synNetCent}A, we show the PageRank scores for these
networks using windows of different length. To use this approach, we
require at least $w$ time steps before we can construct the first
window; this is a potential issue for some
investigations. Additionally, the PageRank scores are sensitive to the
value of $w$. This is another potential difficulty, especially because
one does not know appropriate window lengths in
advance in many situations.  An alternative to aggregating
interactions into discrete sliding windows is to use adjacent windows
of length $w$~\cite{braha2009time}
With this choice, we
aggregate interactions exactly once every $w$ time steps. We show the
resulting time series of PageRank scores in Fig.~\ref{fig:synNetAdjWin} in
Appendix~\ref{sec:synthetic_adjacent_windows}.

When a window includes individual interactions only ($w=1$) or
matches one cycle duration ($w=4$), the PageRank centrality scores capture the
sequence of interactions between the nodes. We show these two cases in
the top row of Fig.~\ref{fig:synNetCent}A. However, even those choices of $w$
are not without issues. Using $w=1$ neglects important temporal
features, and selecting $w=4$ (the cycle length) and placing it
correctly requires prior knowledge of the temporal network, which we
possess only because we invented the rules to create this network.
The extreme sensitivity of PageRank scores to the value of $w$ also causes other problems.
For example, when $w$ is a multiple of the
number of nodes and the period (e.g., $w=20$), this approach fails to
give any useful information at all, because all differences in the
interaction patterns are masked entirely by the window.

\begin{figure}[tp]
  \centering
  \includegraphics[width=.45\textwidth]{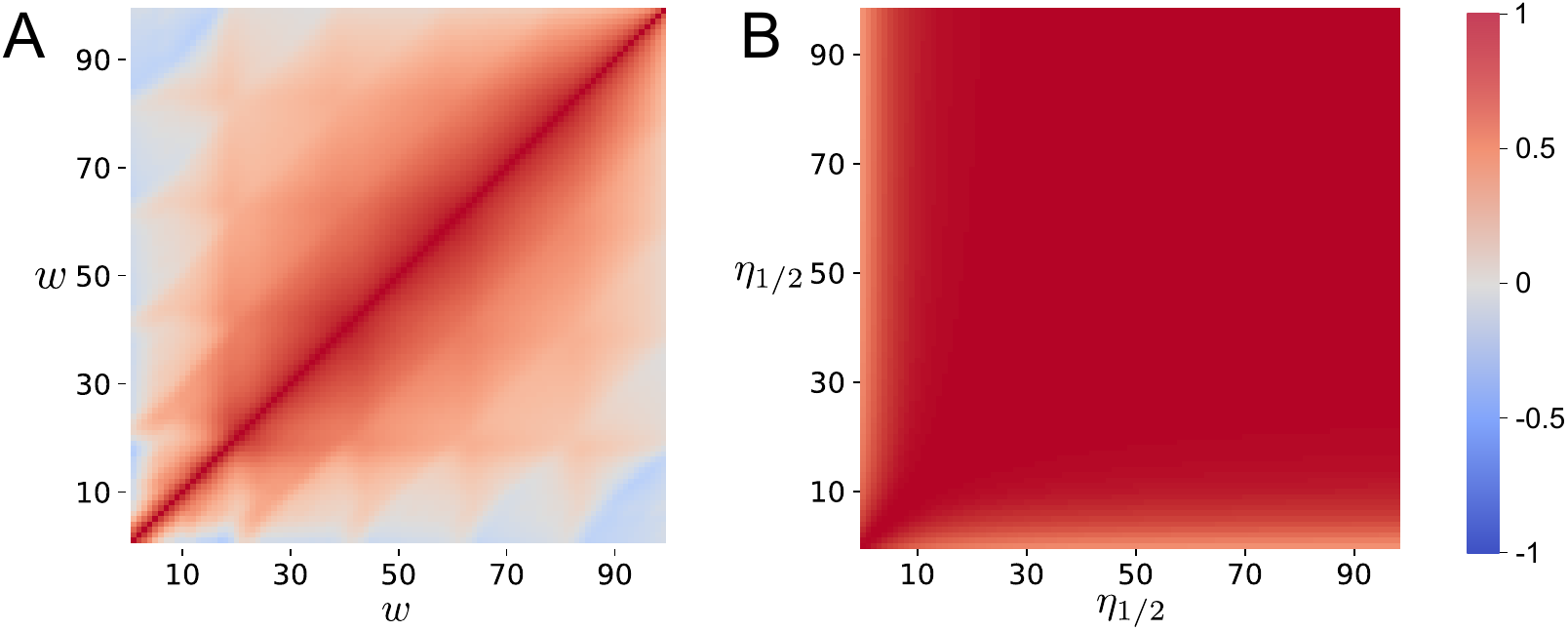}
  \caption{(Color online) {\bf A:} Pearson correlation matrix of the PageRank time series of
    node $1$ in our synthetic network using sliding windows $(t_k - w, t_k]$ for $w \in \{1, \ldots,
    100\}$.  {\bf B:} Pearson correlation matrix of the tie-decay PageRank time series
  of node $1$ for integer half-lives $\eta_{1/2} \in \{1, \ldots, 100\}$.
  }
  \label{fig:corrMatrices}
\end{figure}

We now use our tie-decay formalism to obtain PageRank time series for
different tie half-lives (see Fig.~\ref{fig:synNetCent}B). The
cyclical dynamics of the interactions manifest in all values of the
half-life $\eta_{1/2}$. Different values of $\eta_{1/2}$ entail
different oscillation magnitudes, but the time series includes
oscillations for all choices of $\eta_{1/2}$. Every four time steps,
we observe an increase in the PageRank score of the newly-selected
target node. Using our tie-decay framework also does not have the
problem of `masking' of the dynamics that we illustrated above.

We also examine the sensitivity with respect to the choice of window length and
half-life by computing the Pearson correlation of the PageRank time
series for different values of $w$ and $\tau$ (for ordinary PageRank) and $\eta_{1/2}$ (for tie-decay PageRank). In Fig.~\ref{fig:corrMatrices}A, we show the Pearson correlation between the time series of the (ordinary) PageRank centrality of node $1$ for
each $w \in \{1, 2, \ldots, 100\}$. As the figure shows, binning
interactions over time produces a time series that is very sensitive
to the choice of $w$. The mean correlation between time series is
$0.414$, and the standard deviation is $0.301$. Some time series are
even correlated negatively with each other.  The correlations also
depend on the periodicity of our temporal network. For example,
choices of $w$ for the two time series that differ by exactly $20$
have larger positive correlations with each other than with other
window lengths. By comparison, the Pearson correlations between the
PageRank time series in the tie-decay networks for a variety of values
of the half-life $\eta_{1/2}$ (see Fig.~\ref{fig:corrMatrices}B) paint a very
different picture. We observe large positive
correlations for many pairs of values of $\tau$. The mean correlation
between the time series in $0.945$, which is much larger than what we
observed for our calculation with sliding windows, and the standard
deviation of $0.109$ is much smaller.

Our synthetic example of a temporal network highlights some of the
advantages of our continuous-time tie-decay network formalism for
temporal networks. Results from frameworks that require binning
interactions can be extremely sensitive both to the choice of window
length and to the placement of windows. By contrast, our tie-decay
framework is more robust to parameter choices. Varying $\eta_{1/2}$
(or, equivalently, $\alpha$) adjusts the longevity of ties while
maintaining similar PageRank centrality trajectories, lending
confidence to investigations even when (as is usually the case) one
does not possess precise knowledge of the time scales of the
interactions in a system.


\section{The National Health Service (NHS) Retweet Network}\label{nhs}
We now compute tie-decay 
PageRank scores to track the
evolution of node importances over time in a large data set of
time-annotated interactions on Twitter.

Twitter is a social-media platform that has become a prominent channel
for organizations, individuals, `bots', `sockpuppets', and other types of accounts to broadcast events, share
ideas, report events, and socialize by posting messages (i.e.,
`tweets') of at most 140 characters in
length~\cite{Kwak2010}. (Twitter subsequently expanded the maximum tweet
length to 280 characters, but the maximum was 140 characters at the
time that our data set was collected.) Data from Twitter has allowed
researchers to study patterns and trends 
in a plethora of large-scale political and social events and processes, such as
protests and civil unrest, public health, and information
propagation~\cite{Beguerisse2014, Beguerisse2017, Cihon2016,Giles2012,
  gonzalez2016networked, morales2012users, OSullivan2017,
  tonkin2012twitter}.

Twitter accounts (which can represent an individual, an organization,
a bot, and so on) can interact in several ways, and there are various
ways to encode such interactions in the form of a network. For
example, accounts can subscribe to receive other accounts' tweets (a
`follow' connection), can mention each other in a tweet (a `mention'
connection), can spread a tweet that was posted originally by someone else (a
`retweet' connection) to their followers, and so on. These
interactions represent an explicit declaration of interest from a
source account about a target, and one can thus encode them using
directed networks~\cite{Beguerisse2014}. Because these interactions are 
time-resolved, it is sensible to analyze Twitter networks as time-dependent
networks~\cite{Beguerisse2017}.

\begin{figure}[tp]
  \centering
  \includegraphics[width=0.4\textwidth]{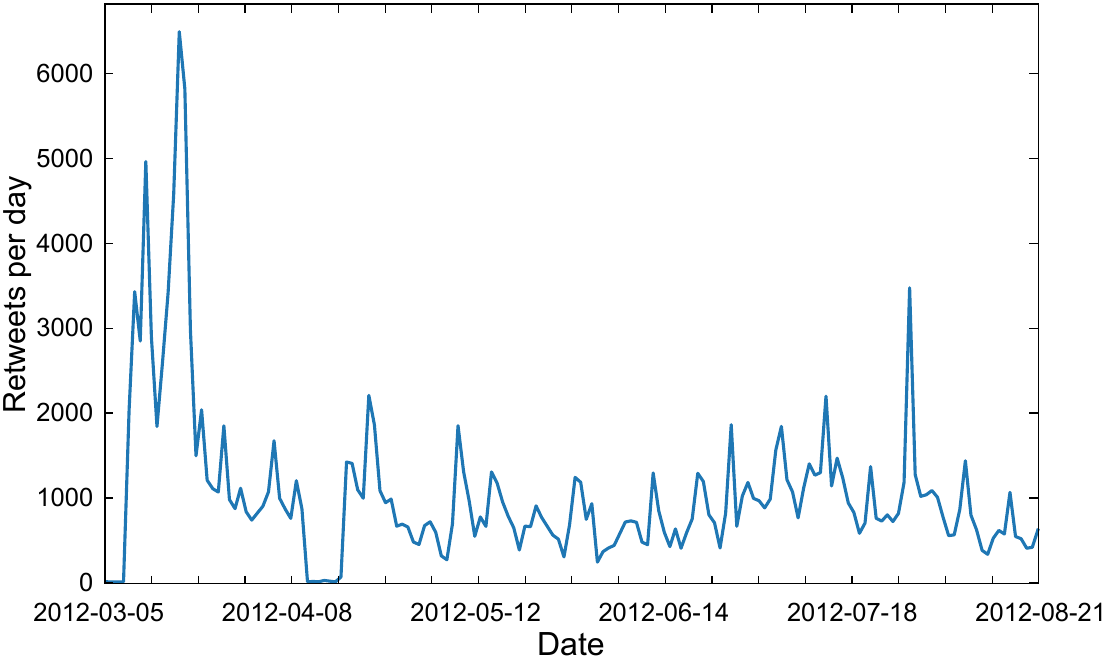}
  \caption[Daily retweet activity in the NHS data set]{Number of daily
    retweets among the 10,000 most-active Twitter accounts in the NHS
    data set.}
  \label{fig:retweetActivity}
\end{figure}

We study a retweet network, which we construct from a data set of
tweets about the United Kingdom's National Health Service (NHS) that
were posted after the controversial Health and Social Care Act of
2012~\cite{healthsocialcare2012}. Our data set covers over five months
of time and includes tweets in English that include the term
`NHS'. Specifically, we consider retweets\footnote{We consider only retweets, so if MAP tweets something and MBD retweets it, then only the retweet can be part of our data set.}
 by the 10,000 most-active Twitter accounts
(according to the number of tweets in our data set) from 5 March
2012 through 21 August 2012 (see Fig.~\ref{fig:retweetActivity}). All
data were collected by Sinnia\footnote{See
  \url{http://www.sinnia.com/}.}, a data-analytics company, using
Twitter Gnip PowerTrack API\footnote{See
  \url{https://gnip.com/realtime/powertrack/}.}. From these data, we
construct a tie-decay temporal network in which the interactions are
retweets\footnote{In our computations, we set ties $b_{ij} < 10^{-7}$
  to $0$ to preserve the sparsity of $B(t)$. Consequently, nodes may
  rejoin the dangling-node set.}.
Our code for constructing tie-decay networks is available at \url{https://bitbucket.org/walid0925/tiedecay/}.


\subsection{Tie-Decay PageRank Centrality in the NHS Retweet Network}
The temporal network of retweets from the NHS data has the
tie-strength matrix $\matt{B}$ (see equation~\eqref{eq:Bupdate}) and
starts from the initial condition $\mat{B}(0)=0$. We construct three
tie-decay networks: ones whose values of $\alpha$ correspond to tie
half-lives of $1$ hour, $1$ day, and $1$ week. We compute tie-decay
PageRank scores of all Twitter accounts for each of these networks.

\begin{figure}
  \centering
   \includegraphics[width=.45\textwidth]{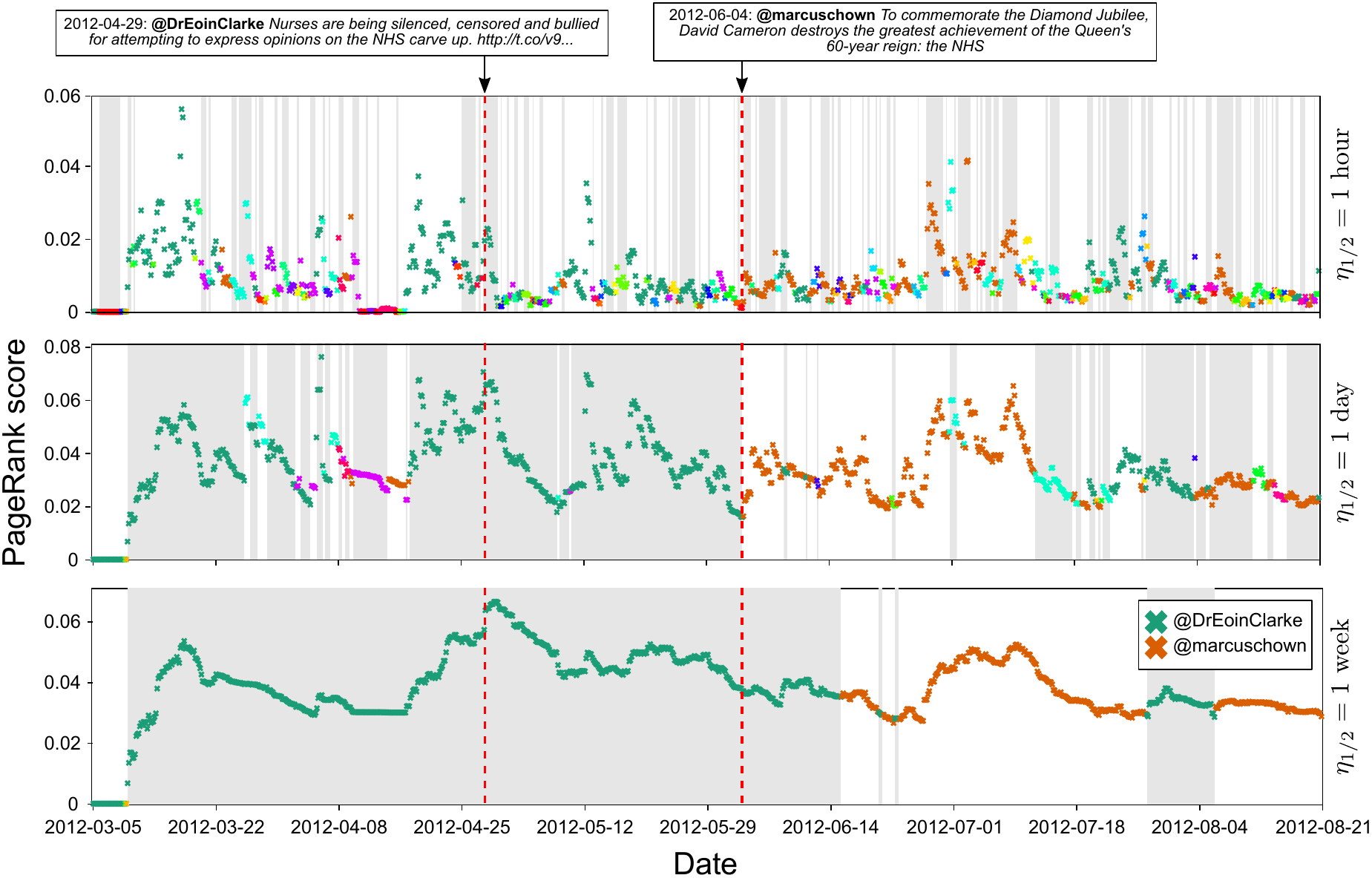}
   \caption{(Color online) The top Twitter accounts, according to
     tie-decay PageRank, in the temporal NHS retweet network with
     three values of tie half-life. Each 
     color is associated with
     a unique Twitter account, and the alternating gray and white
     background color indicates intervals in which the same account
     has the top tie-decay PageRank score. Transitions in color
     (from white to gray, and vice versa) indicate when there is a change in which account has the top score.
     The first dashed red vertical line
     indicates the time
     that {\sc @DrEoinClarke} posted the tweet in
     the associated box; the second such line corresponds to a tweet by
     {\sc @marcuschown}.  }
  \label{fig:decayPRalphas}
\end{figure}

In Fig.~\ref{fig:decayPRalphas}, we show an example of the effect of
the value of $\alpha$ on PageRank scores. We compute tie-decay PageRank
scores for networks in which the tie half-life is $1$ hour, $1$ day,
and $1$ week. In each panel of Fig.~\ref{fig:decayPRalphas}, we plot
the Twitter account with the largest PageRank score at every time
point; the transitions between white and gray shading indicate when there is a change in which account has the top score.
 When the
half-life is short (i.e., $\alpha$ is large), interactions produce
feeble ties that die off quickly unless there are frequent and
sustained interactions between the accounts. Consequently, the
tie-decay PageRank scores of the Twitter accounts change wildly in time, and (as
illustrated in the top panel in Fig.~\ref{fig:decayPRalphas}) such a
short half-life implies that the Twitter account with the top tie-decay PageRank
score changes frequently. When the half-life is longer (e.g., $1$
day), ties are better able to build momentum and strengthen from
interactions that otherwise would be too far apart in time. The ability to build and maintain ties results in fewer transitions between which
Twitter account has the top spot in the ranking. The middle panel in
Fig.~\ref{fig:decayPRalphas} illustrates that we indeed observe
less-frequent transitions when the half-life is $1$ day.  Finally,
when the half-life of a tie is $1$ week, two specific accounts ({\sc
  @DrEoinClarke} and {\sc @marcuschown}) dominate the ranking; they
alternate between the top and second spots.

In this case study, we reveal two accounts as dominant ones as we tune
the half-life of ties to larger values. The first of these, Eoin
Clarke ({\sc @DrEoinClarke}), is a Labour-party activist and was an
outspoken critic of the UK coalition government's stance in 2012 on
the NHS. Marcus Chown ({\sc @marcuschown}) is a science writer,
journalist, and broadcaster who was also an outspoken critic of the UK
government's NHS policy in 2012. Their dominance becomes apparent
as we increase the half-life of the ties. On 29 April, {\sc
  @DrEoinClarke} posted a tweet that gathered significant
attention. This tweet yields a short-lived boost in his PageRank score
when $\eta_{1/2}$ is $1$ hour and $1$ day, and it yields a more
sustained increase when $\eta_{1/2}$ is $1$ week. On 4 June, {\sc
  @marcuschown} posted a tweet that boosts
  his PageRank score. When $\eta_{1/2}$ is $1$ day, the number of retweets of this
tweet are enough to carry him to the top spot. However, when
$\eta_{1/2}$ is $1$ week, the retweets are not enough to overtake {\sc
  @DrEoinClarke}, whose ties remain strong.

\begin{figure}[tp]
  \centering
  \includegraphics[width=.45\textwidth]{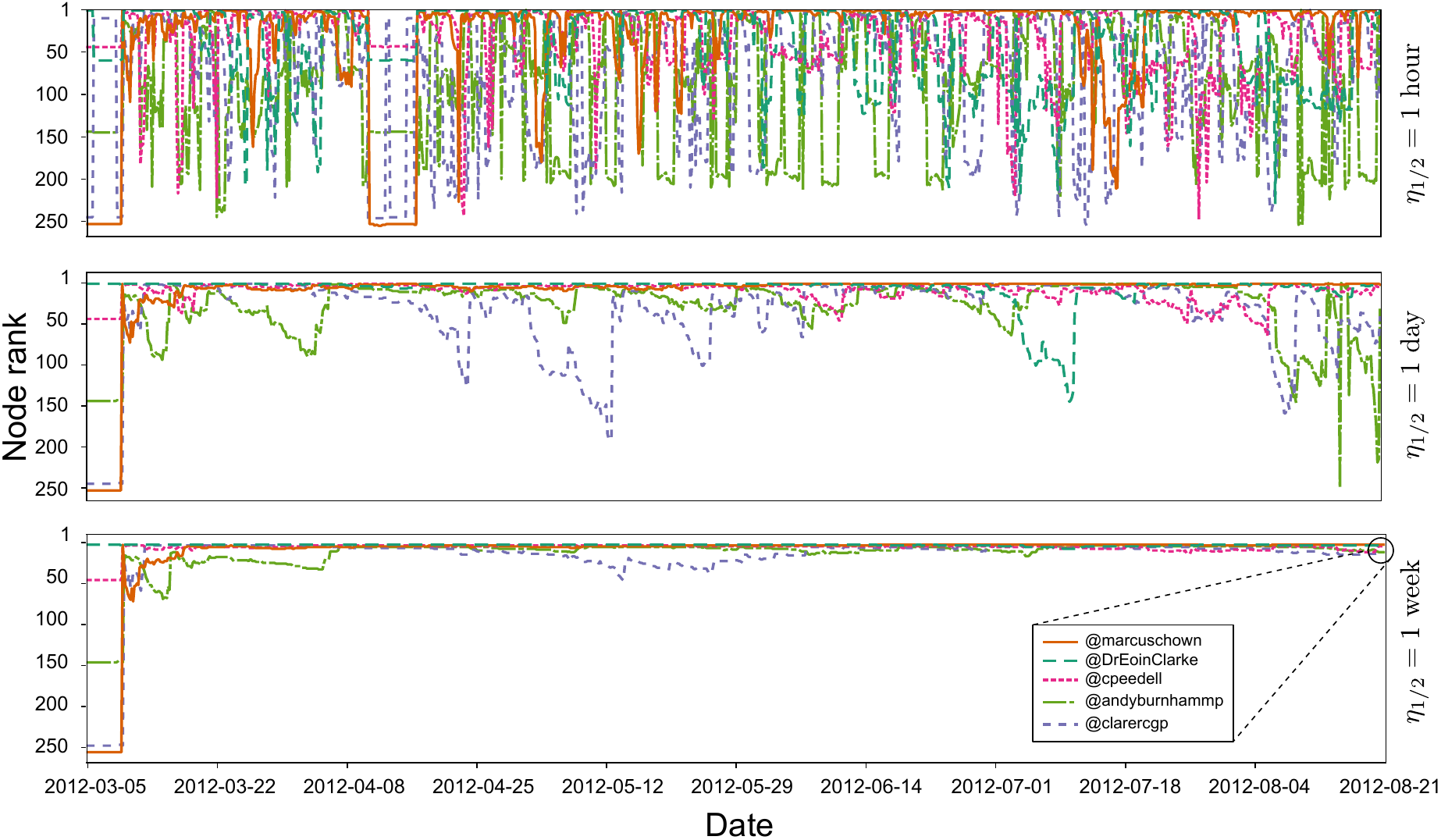}
  \caption{(Color online) Time series of 
  time-resolved ranks of five prominent Twitter accounts for aggregations of tie-decay networks
    with three different values of the half-life. More-important
    accounts are higher on the vertical axis.}
  \label{fig:ranking_switch}
\end{figure}

In Fig.~\ref{fig:ranking_switch}, we show a complementary illustration
of the effect of half-life value on the time-resolved PageRank rankings of
Twitter accounts. We construct a time-independent network in which we
aggregate all of the interactions in our data set --- specifically, we
consider $B(t)$ for $\alpha=0$ with the time $t$ set to be 21 August
2012 --- and we determine the top-5 accounts by calculating the
standard time-independent version of PageRank (with $\lambda=0.85$) on
this network. We then track the 
time-dependent 
PageRank ranks (where rank $1$
is the Twitter account with the largest tie-decay PageRank score, and so on) of
these five accounts for different values of $\alpha$ (or,
equivalently, of $\eta_{1/2}$). When $\eta_{1/2}$ is $1$ hour, these
accounts often overtake each other in the rankings, and the changes in
rankings can be rather drastic, as some Twitter accounts drop or rise
by almost $250$ spots. As we consider progressively longer
half-lives, we observe less volatility in the rankings.

The experiments in Figs.~\ref{fig:decayPRalphas}
and~\ref{fig:ranking_switch} demonstrate how one can use $\alpha$ as a
tuning parameter to reflect the longevity of relationship values in a
temporal network. They also demonstrate the value of our tie-decay
formalism for illustrating fluctuations in network structures. When
analyzing networks in discrete time, there is a risk that aggregating
interactions may conceal intermediate dynamics and nuances of network
structure~\cite{fenn2012}. By contrast, our continuous-time network
formalism avoids arbitrary cutoff choices (and potential ensuing
biases \cite{kivela2015}) when choosing the borders of time windows.
It also allows a smoother exploration of network structure at a level of
temporal granularity that is encoded in the value of $\alpha$.

\subsection{Aggregating Interactions versus Examining Tie-Decay Networks}
\label{sec:aggvdecay}
\begin{figure}[tp]
  \centering
  \includegraphics[width=.5\textwidth]{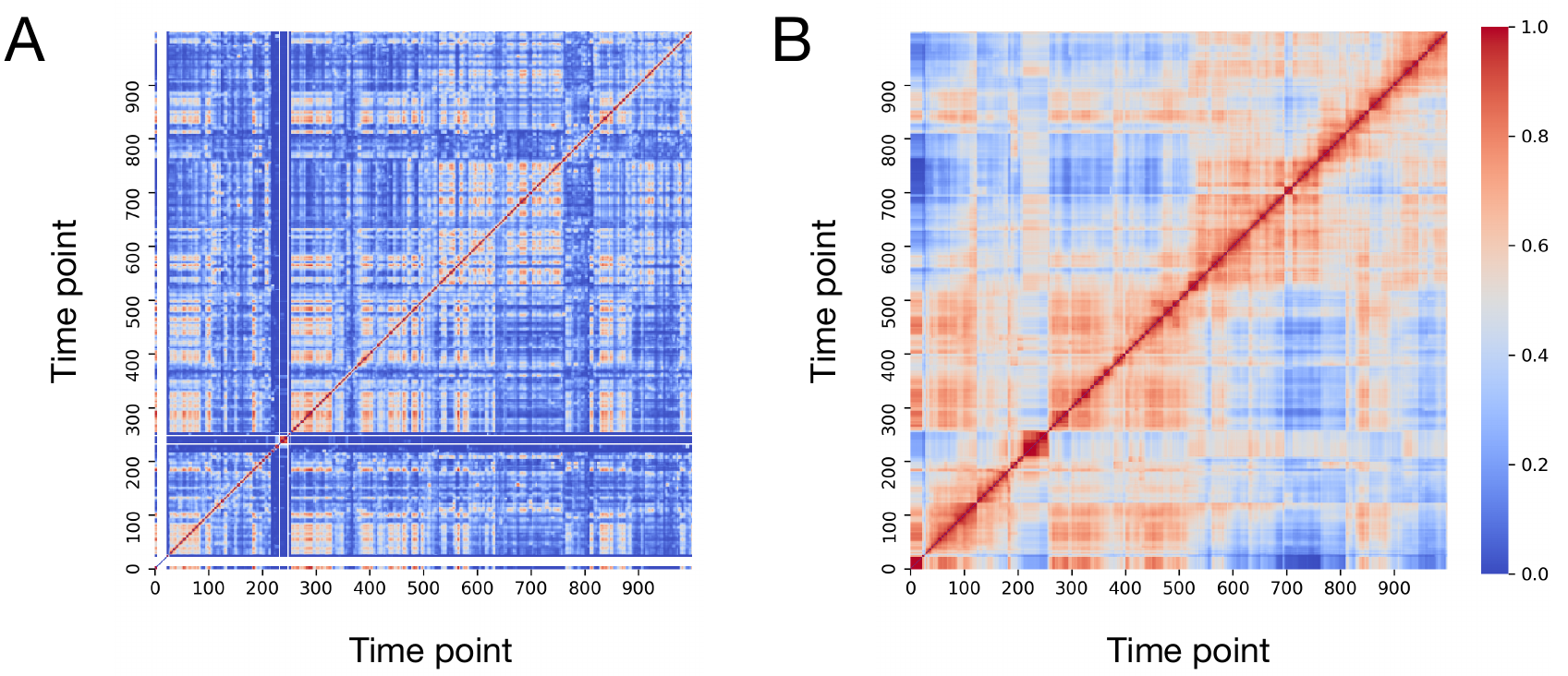}
  \caption{(Color online) {\bf A:} Pearson
    correlations between PageRank vectors in networks that we
    construct using sliding windows of $w=1~\text{day}$. 
    {\bf B:} Pearson correlations in tie-decay
	networks with $\eta_{1/2} = 1~\text{ day}$. 
    }
  \label{fig:NHSCorrMatrices}
\end{figure}

Using tie-decay networks instead of aggregating interactions via
sliding windows can have a large impact on the qualitative
results of an investigation. To illustrate this using the NHS data,
we compare
{the Pearson correlation between the PageRank vectors}
from tie-decay networks to the Pearson correlation on networks that we construct using sliding windows.
 We sample $K$
equally-spaced time points between the first and last interactions in
the data.\footnote{{The first interaction occurs at
    20:41:46 on 5 March 2012, and the last interaction occurs at 09:09:25 on 21 August
    2012.}} At each time point $t_k$, we construct a time-independent
network with interactions in the time window $(t_k - w, t_k]$ for a
  given window length $w$, and we compute the PageRank vector for this network.
  In Figure~\ref{fig:NHSCorrMatrices}A, we show the Pearson correlation
  matrices between PageRank vectors using $K=1,000$ and $w = 1$
  day. The sampled time points are approximately four hours apart, so
  consecutive time points have overlapping windows. However,
  the Pearson correlations between the PageRank vectors from the different time-independent networks
  are relatively small. This is usually the case even for the PageRank vectors from time points that are near each other.
    Such small temporal correlations are not surprising;
  previous research~\cite{braha2006centrality} has reported that the
  connections in networks that one constructs by aggregating interactions on
  different days can vary drastically. When this occurs, the most
  central nodes in networks from different days can also differ
  drastically. 

We also calculate the tie-decay PageRank vectors using the same time
points. In this case, we observe a pronounced ``block structure" in
the correlation matrix (see Fig.~\ref{fig:NHSCorrMatrices}B). That is, PageRank
vectors from time points that are close to each other have large
correlations with each other because tie strengths of past
interactions persist in time. We can adjust the extent of such
persistence by varying the half-life. In Figure~\ref{fig:NHSCorrMatDists}
in Appendix~\ref{sec:appAggvDecay}, we show the distributions of
pairwise correlation values using both approaches.

\subsection{Computational Efficiency}
\label{sec:efficiency}

\begin{figure}[tp]
  \centerline{\includegraphics[width=.45\textwidth]{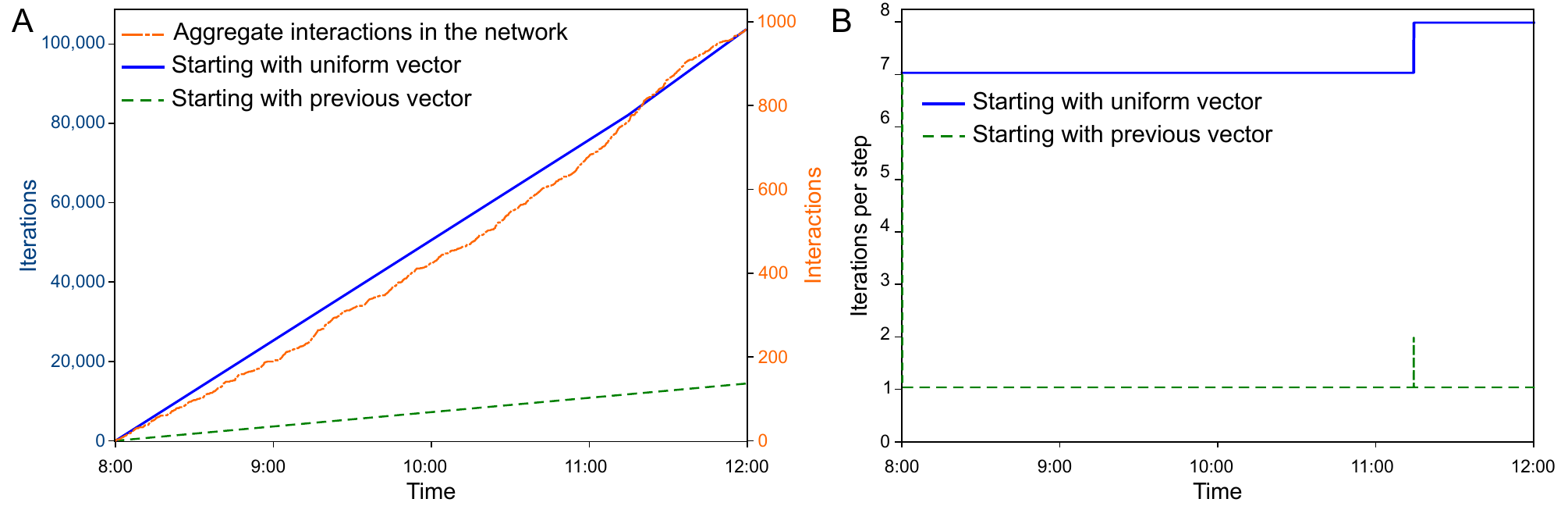}}
  \caption{(Color online) {\bf A:} Cumulative number of iterations to
    convergence (where we define `convergence' as
    $\|\vpi^{\left(k+1\right)}(\tdt) -
    \vpi^{\left(k\right)}(\tdt)\|_{l_1} <10^{-6}$) for calculating tie-decay PageRank on the NHS tweet network when the starting
    vector is the uniform vector (solid blue curve) and the previous
    PageRank vector (dashed green curve). For context, we also include
    the aggregate number of interactions in the network
    (dash--dotted orange curve) and label it on the right vertical axis. {\bf B:} When
    analyzing the NHS network, we observe that tie-decay PageRank requires at
    most $2$ iterations to converge when we start from the previous
    time step's PageRank vector, whereas using the uniform vector requires $7$
    or more iterations. In this example, the half-life of a tie is $1$
    day.  
    }
  \label{fig:prsconv}
\end{figure}

In applications (e.g., for streaming data), it is often desirable to
update the values of time-dependent centrality measures, such as
tie-decay PageRank scores, each time that there is a new interaction. We know
from Theorem~\ref{thm:boundPR} that there is a bound on the magnitude
of the difference between the PageRank vectors at times $t$ and $\tdt$
when there is a new interaction. Consequently, we expect to obtain
faster convergence of the iteration for $\tdt$ when we use the
PageRank vector from $t$ as our initial vector than if we start the computation from scratch. 
To demonstrate this, we select a period of time with high activity in our NHS data ---
08:00 am to 12:00 pm on 18 March 2012 (see
Fig.~\ref{fig:retweetActivity}) --- and calculate the tie-decay
PageRank vector at time $\tdt$ with two different starting vectors:
the uniform vector $\vpi^{\left(0\right)}(\tdt)=\frac{1}{n}\mathbb{1}$ and the previous
PageRank vector $\vpi^{\left(0\right)}(\tdt)=\vpi(t)$. In
time-independent networks, it has been observed that the uniform
vector has better convergence properties than any other starting
vector, in the absence of prior knowledge about the final PageRank
vector~\cite{Gleich2015}. In our tie-decay network formalism, given
the bound between the magnitudes of the PageRank vectors at times $t$ and $\tdt$, it
is intuitive that using the vector from the previous time has
computational advantages over other choices. We demonstrate this fact
in Fig.~\ref{fig:prsconv}.

\section{Conclusions and Discussion}\label{conclude}

We have introduced a continuous-time framework that incorporates tie
decays for studying temporal networks, and we used our formalism ---
which we call `tie-decay networks' --- to generalize PageRank
centrality to continuous time.

In our proposed tie-decay formalism, a tie between two nodes
strengthens through repeated interactions and it decays in their
absence. Such tie-decay networks allow one to tractably analyze
time-dependent interactions without having to aggregate interactions
into time windows (i.e., bins), as is typically done in existing frameworks for
studying temporal networks~\cite{holme2015modern,holme2019}. We purposely
avoided aggregating interactions using time bins, whose sizes and
placement are difficult to determine, by modeling the weakening of
ties in time as exponentially decaying with a rate of $\alpha$ (or,
equivalently, with a half-life of $\eta_{1/2}$). In addition to
representing the decay of human relationships~\cite{Burt2000}, as we
have done in this paper, it is also possible to use our formalism
more generally to model the decreasing value of old information, decay in
other types of interactions, and so on.

We showcased our tie-decay formalism on both a synthetic temporal network and a
network of retweets on Twitter. 
Our computations illustrated that
adjusting the value of the half-life $\eta_{1/2}$
allows one to examine the
temporal dynamics of rankings at different time scales of
interest. Our examination of a synthetic network illustrated that using
tie-decay networks can help mitigate serious issues that can arise form binning interactions, such as
sensitivity to time scales and the masking of interaction
patterns. To provide a case study for the study of tie-decay PageRank
centralities in an empirical temporal network, we investigated the
temporal evolution of the ranks of important accounts in a large collection of
retweets about the UK's National Health Service. 
We also developed a numerical scheme and bounds on the change of tie-decay PageRank scores upon the arrival of each new
interaction. Such bounds are important for studying data streams, in
which new data arrives at a potentially alarming rate. By tuning the
decay rate of interactions, we illustrated that PageRank scores
can change much more drastically when the half-life is short than when the
half-life is long.

Our tie-decay formalism for continuous-time changes in network
architecture provides an important step for the study of streaming
network data and the development of tools to analyze temporal networks
in real time. Streaming data is ubiquitous --- it arises in social-media
data, sensor streams, communication networks, and
more~\cite{Gaber:miningdatastreams,Latapy2017} --- and analyzing
tie-decay networks offers a promising approach for studying it.
In the present paper, we illustrated how to perform an update of tie-decay PageRank from one
time step to another, and it will be important to develop such ideas
further for other types of computations (such as community detection and other types of clustering).
In the short term, it will be useful to implement efficient schemes for numerical computations of tie-decay
generalizations of other centralities scores (such as hubs and
authorities) that are defined from eigenvectors.  For instance, our
framework permits a tie-decay generalization of personalized
PageRank~\cite{Gleich2015,jeub2015} (by making a different choice of $\vecb{v}$ in equation~\eqref{eq:priteration}), which one can use in turn to develop
new, principled methods for studying local community structure in networks
that evolve in continuous time. Our tie-decay formalism is also very
well-suited to incorporating time-dependent strategies for
teleportation in PageRank~\cite{gleich-dynamic}. An extension of our tie-decay
formalism to noninstantaneous interactions (i.e., taking durations
into account) is possible by replacing the term with the Dirac
$\delta$ function in equation~\eqref{eq:tie_ode} with a function that
is nonzero only when a tie exists. For example, if an interaction
lasts from $\tau_{\mathrm{begin}}$ until $\tau_{\mathrm{end}}$, then
$H(t - \tau_{\mathrm{begin}}) - H(t - \tau_{\mathrm{end}})$, where $H$
is the Heaviside step function. One can also formulate interactions
with time durations using window functions~\cite{Weisstein2002} or test
functions~\cite{Howison2003}. 

A wealth of other avenues are also worth pursuing. For example, it is desirable to 
systematically investigate heterogeneous
decay rates (e.g., individual
rates for nodes or ties), fit the decay parameter to data, use
decay functions other than exponential ones~\cite{gleeson2014,
  yang2018},\footnote{It may be particularly interesting to explore
  the effects of heavy tails in decay rates. As explained
  in~\cite{Masuda2018}, one can express a power law as a mixture of
  exponentials, facilitating the incorporation of heavy tails into our tie-decay formalism.}, develop clustering methods for tie-decay networks,
analyze localization phenomena (and their impact on centralities
and clustering)~\cite{martin2014,Taylor2015}, develop and study
random-network null models for tie-decay networks, incorporate
noninstantaneous interactions, investigate change-point detection, and examine
continuous-time networks with multiplex interactions. It is also desirable to study a variety of dynamical processes --- such as contagion spread and opinion dynamics --- on tie-decay networks.

Many empirical networks and data sets from which one can construct networks
are time-dependent, and it is important to be
able to model such systems in continuous time. Tie-decay networks
offer a promising approach for further development of continuous-time
temporal networks.


\section*{Acknowledgements}

We thank Mauricio Barahona, Alain Goriely, Peter Grindrod, Mikko
Kivel\"a, Renaud Lambiotte, Naoki Masuda, Fabian Ying, and Xinzhe Zuo
for useful discussions. We thank Guillermo Gardu\~no and Sinnia for
their help in collecting the data. MBD acknowledges support from the Oxford--Emirates Data Science Laboratory.


\appendix

\section{Proofs} \label{sec:proof_boundPR}


\subsection{Proof of Lemma~\ref{lemma:no_inter}}

\begin{proof}
When there is a new interaction from node $i$ to
node $j$, the rows of $P(\tdt)$ that correspond to nodes
$k \neq i$ (i.e., nodes from which the new connection does not
originate) are unchanged: $p_{kh}(\tdt) = p_{kh}(t)$ for all
$h\in\{1,\ldots, n\}$.  

To determine the change in the $i$-th row of
$P(t + \Delta t)$, we first examine the $i$-th row of
$B(t + \Delta t)$ by calculating
\begin{align}
  b_{ih}(t + \Delta t) =
  \begin{cases}
    e^{-\alpha \Delta t}b_{ih}(t)\,,\qquad & h \neq j \\
    e^{-\alpha \Delta t}b_{ij}(t) + 1\,, \qquad & h = j\,.
  \end{cases}
\end{align}

We then consider the change to the rank-$1$ correction 
$\vecb{c}(t)\vecb{v}^T$. If $i$ is not a dangling node at time $t$,
then $\vecb{c}(t+\Delta t) = \vecb{c}(t)$. However, if $i$ is a
dangling node at time $t$, then $ c_i(t) = 1$ and $c_i(t+\Delta t) =
0$.  Therefore,
\begin{align}
  c_i(t+\Delta t) - c_i(t) = 
  \begin{cases}
    0\,, \quad & c_i(t) = 0 \\
    -1\,,  \qquad & c_i(t) = 1\,.
  \end{cases}
\end{align}
Observe that $c_i(t+\Delta t)$ necessarily equals $0$ and that $c_i(t)
\in \lbrace 0 , 1 \rbrace$. Therefore,
\begin{equation}
  c_i(t+\Delta t) - c_i(t) = -c_i(t)\,,
\end{equation}
so the change to the correction term is
\begin{equation}
  c_{i}(t+\Delta t)v_i - c_{i}(t)v_i = -c_{i}(t)v_i\,.
\end{equation}

The $i$-th row of $P(t + \Delta t)$ is
\begin{align}
  p_{ih}(\tdt) =
  \begin{cases}
    \dfrac{e^{-\alpha \Delta t}b_{ih}(t)}{1+e^{-\alpha \Delta t}\sum_k
      b_{ik}(t)} - c_{i}(t)v_i\,, \qquad & h \neq j \\
    \\
    \dfrac{e^{-\alpha \Delta t}b_{ij}(t)+1}{1+e^{-\alpha \Delta
        t}\sum_k b_{ik}(t)} - c_{i}(t)v_i\,, \qquad & h = j\,.
  \end{cases}
\end{align}
For $h \neq j$, the difference between $p_{ih}(\tdt)$ and $p_{ih}(t)$ is
{\tiny
\begin{align}
  p_{ih}(\tdt) - p_{ih}(t) &= \frac{e^{-\alpha \Delta t}b_{ih}(t)}
  {1+e^{-\alpha \Delta t}\sum_k b_{ik}(t)} - \frac{b_{ih}(t)}{\sum_k
    b_{ik}(t)} - c_i(t)v_i  \notag \\
  &= \frac{-b_{ih}(t)}{\sum_k
    b_{ik}(t)\left[1 + e^{-\alpha \Delta t}\sum_k b_{ik}(t)\right]} -
  c_i(t)v_i \notag \\
  &= \frac{-b_{ih}(t)}{d_{ii}(t)\left[1+e^{-\alpha \Delta
        t}d_{ii}(t)\right]} - c_i(t)v_i\,.
\end{align}}
When $h = j$, we have
{\tiny
\begin{align}
  p_{ij}(t+ \Delta t) - p_{ij}(t) &= \frac{1+e^{-\alpha \Delta
      t}b_{ij}(t)} {1+e^{-\alpha \Delta t}\sum_k b_{ik}(t)} -
  \frac{b_{ij}(t)}{\sum_k b_{ik}(t)} - c_i(t)v_i \notag \\
  &= \frac{\sum_k
    b_{ik}(t)-b_{ij}(t)} {\sum_k b_{ik}(t)\left[1+ e^{-\alpha \Delta
        t}\sum_k b_{ik}(t)\right]} - c_i(t)v_i \notag \\
  &= \frac{d_{ii}(t)-b_{ij}(t)}{d_{ii}(t)\left[1+ e^{-\alpha \Delta
        t}d_{ii}(t)\right]} - c_i(t)v_i  \notag \\
  &= \frac{1}{1+e^{-\alpha
      \Delta t}d_{ii}(t)}-\frac{b_{ij}(t)}
       {d_{ii}(t)\left[1+e^{-\alpha \Delta t}d_{ii}(t)\right]} -
       c_i(t)v_i\,.
\end{align}}
In matrix terms, the change from $P(t)$ to $P(\tdt)$ is thus
\begin{align}
  \dP  &= \frac{1}{1+e^{-\alpha \Delta t}d_{ii}(t)}\vecb{e}_{i}\vecb{e}_{j}^T \notag \\
 &\quad  - \frac{1}
  {d_{ii}(t)\left(1+e^{-\alpha \Delta t}d_{ii}(t)\right)}
  \vecb{e}_{i}\vecb{e}_{i}^T B(t) -c_i(t)v_i\vecb{e}_i\mathbb{1}^T\,,
\end{align}
which concludes the proof.
\end{proof}


\subsection{Proof of Theorem~\ref{thm:boundPR}}

\begin{proof}
The change in PageRank scores with one new interaction is
{\tiny
\begin{align}
 	   \vpi(\tdt) - \vpi(t) &= 
     \left[\lambda(P(t)^T + \dP^T) 
       + (1-\lambda)\vecb{v}\mathbb{1}^T\right]\vpi(\tdt)  \notag \\
  &\quad - \left[\lambda P(t)^T 
      + (1-\lambda)\vecb{v}\mathbb{1}^T\right]\vpi(t)  \\
    & = \lambda P(t)^T(\vpi(\tdt) - \vpi(t)) 
      + \lambda \dP^T\vpi(\tdt)\,. \notag
\end{align}}
Rearranging terms gives
\begin{equation}
  \left(I_n - \lambda P(t)^T\right)\left(\vpi(\tdt) 
  - \vpi(t)\right)  
  = \lambda \Delta {P}^T \vpi(\tdt)\,,
\end{equation}
which implies that
\begin{equation} \label{eq:vpi_diff}
  \vpi(\tdt) - \vpi(t) = 
  \lambda  \left(I_n - \lambda P(t)^T\right)^{-1}\dP^T
  \vpi(\tdt)\,,
\end{equation}
where $I_n$ is the $n \times n$ identity matrix. From a Neumann-series expansion~\cite{tyrtyshnikov_1997}, we see that $\norm{(I_n - \lambda P(t)^T)^{-1}}_1$ is bounded above by $1/(1 - \lambda)$.

Taking norms on both sides of~\eqref{eq:vpi_diff} yields
\begin{equation}
	  \norm{\vpi(\tdt) - \vpi(t)}_{1} 
  \leq \frac{\lambda}{1-\lambda}\norm{\Delta P^T}_{1}\,.
  \label{eq:prdiffbound}
\end{equation}
Noting that $\norm{\dP^T}_{1} = \norm{\dP}_\infty$, 
we use the definition of $\dP$ from equation~\eqref{eq:deltaP} to obtain
{\tiny
\begin{equation}
  \begin{split}
    \norm{\vpi(\tdt) - \vpi(t)}_{1} \leq &
    \frac{\lambda}{(1-\lambda)(1+e^{-\alpha \Delta t}d_{ii}(t))}
    \norm{ \vecb{e}_{i}\vecb{e}_{j}^T -\frac{1}{d_{ii}(t)}
      \vecb{e}_{i}\vecb{e}_{i}^TB(t)}_{\infty} \\ & - \frac{\lambda
      c_i(t)v_i}{1-\lambda}\norm{
      \vecb{e}_{i}\mathbb{1}^T}_\infty\,.
  \label{eq:prdiffinserted}
\end{split}
\end{equation}}

Recall that $B(t)$ is the tie-strength matrix and that $\vecb{e}_{i}$
and $\vecb{e}_{j}$, respectively, are the $i$-th and $j$-th canonical
vectors. Let $Q = \vecb{e}_{i}\vecb{e}_{j}^T -\frac{1}{d_{ii}(t)}
\vecb{e}_{i}\vecb{e}_{i}^TB(t)$ be the matrix with elements
\begin{align}
	q_{hk} = 
\begin{cases}
  1 - b_{hk}(t)/d_{ii}(t) \,, & \qquad h = i\,, k = j \\
  - b_{hk}(t)/d_{ii}(t) \,, & \qquad h = i\,, k \neq j  \\
  0 \,, & \qquad \text{otherwise}\,,
\end{cases}
\end{align}
so $Q$ has nonzero elements only in row $i$. 
Noting that $d_{ii}(t) = \sum_{k} b_{ik}(t)$ and using
\begin{equation}
	\norm{Q}_\infty = \max_{1 \leq h \leq n} \sum_{k=1}^n |q_{hk}|
        = \sum_{k=1}^n |q_{ik}|\,,
\end{equation}
we see that
\begin{equation}
\label{qinf_norm}
	\norm{Q}_\infty \leq 2\,.
\end{equation}

We also observe that $\vecb{e}_i \mathbb{1}^T$ is the $n \times n$ matrix whose elements are equal to $1$ in row $i$ and are equal to $0$ elsewhere. 
Therefore, 
\begin{equation}
\label{e1inf_norm}
	\norm{\vecb{e}_i \mathbb{1}^T}_\infty = n\,.
\end{equation}
With $nv_i = 1$ (i.e., uniform teleportation), it follows from equations~\eqref{eq:prdiffinserted},~\eqref{qinf_norm}, and~\eqref{e1inf_norm} that
\begin{equation}
  \norm{\vpi(\tdt) - \vpi(t)}_{1} \leq
  \frac{2\lambda}{(1-\lambda)(1+e^{-\alpha \Delta t}d_{ii}(t))} -
  \frac{\lambda c_i(t)}{1-\lambda}\,.
  \label{eq:upperprboundspecific}
\end{equation}

The change in the PageRank vector is also subject to the
bound~\cite{Lee2003,Ng2001}
\begin{equation}
  \norm{\vpi(\tdt) - \vpi(t)}_1 \leq 
  \frac{2\lambda}{1-\lambda} \sum_{s \in \mathcal{S}(\tdt)}\pi_{s}(t)
  = \frac{2 \lambda }{1-\lambda}\pi_{i}(t)\,,
  \label{eq:prmarkovbound}
\end{equation}
where $\mathcal{S}(\tdt)$ is the set of nodes (in this case, just node
$i$) that experience a change in transition probabilities (i.e., a
change in out-edges). Combining the bounds in \eqref{eq:upperprboundspecific} and \eqref{eq:prmarkovbound} yields
{\tiny
\begin{equation}\label{ineq}
	  \norm{\vpi(\tdt) - \vpi(t)}_1 
  \leq \frac{2\lambda}{1-\lambda} \min 
  \left\{ \pi_{i}(t), \, 
    \frac{1}{1+e^{-\alpha \Delta t}d_{ii}(t)} - \frac{c_i(t)}{2} \right\}\,,
\end{equation}}
which completes our proof.
\end{proof}


Note that $d_{ii}(t) > 0 \iff c_i(t) = 0$ and $d_{ii}(t) = 0 \iff
c_i(t) = 1$, which guarantees that the quantity on the right-hand side
of \eqref{ineq} is always positive. This gives the results in
Corollaries~\ref{corr:dangling} and \ref{corr:notdangling}.

\subsection{Synthetic Network: Using Adjacent Time Windows}
\label{sec:synthetic_adjacent_windows}
\begin{figure}[htp]
  \centering
  \includegraphics[width=.45\textwidth]{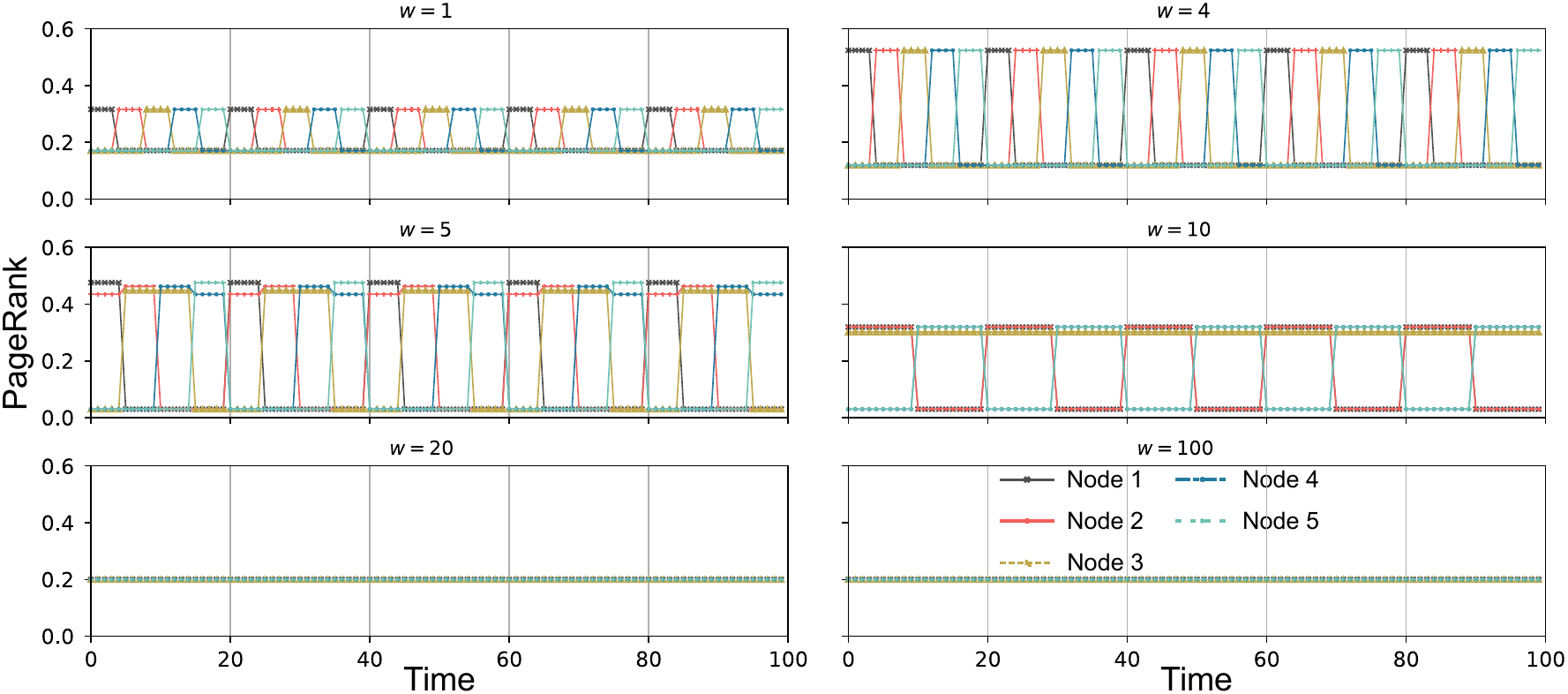}
  \caption{(Color online) 
  PageRank scores over time for each node in our synthetic example
    using adjacent time windows.}
  \label{fig:synNetAdjWin}
\end{figure}

In our synthetic network from Section~\ref{sec:synthetic}, suppose that we aggregate the interactions into adjacent (i.e., non-overlapping) windows of length $w$. We then calculate PageRank time series for the resulting sequence of time-independent networks and show our results in Fig.~\ref{fig:synNetAdjWin}.

\subsection{NHS Retweet Network: Network Statistics}
\label{sec:NHSnetworkstats}|

\begin{figure}[htp]
  \centering
  \includegraphics[width=.45\textwidth]{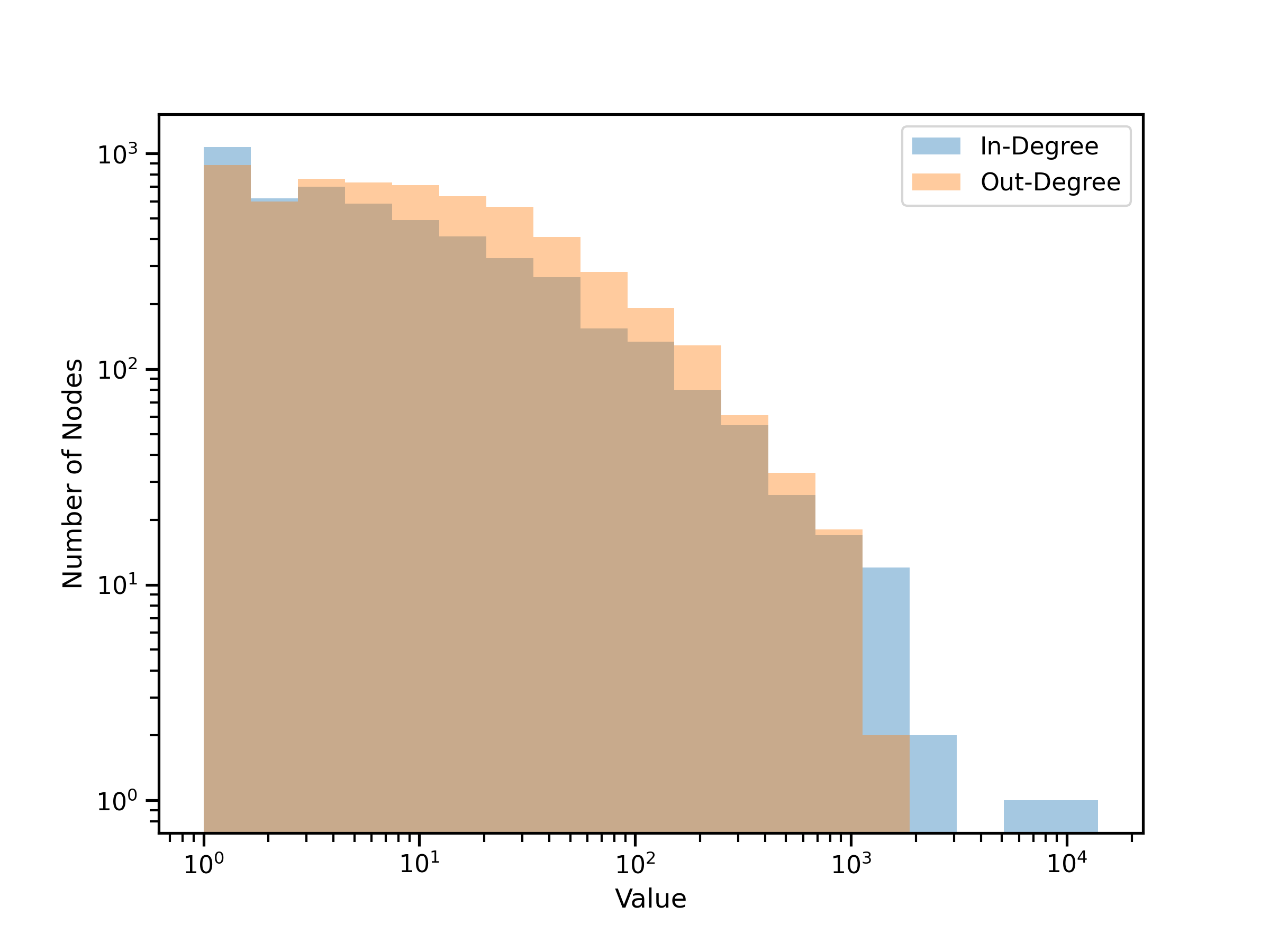}
  \caption{(Color online) 
  Distributions of retweets received (i.e., in-degree) and sent (i.e., out-degree) in the NHS retweet network between 5 March 2012 and 21 August 2012.
  }
  \label{fig:retweetsUserDist}
\end{figure}

The NHS network includes retweets between the 10,000 most-active accounts, where we measure their activity as their number of tweets in the data set. 
   There are 181,123 retweets between the 10,000 most active accounts between 5 March 2012 and 21 August 2012.
   We interpret each of these retweets as one interaction.
 In total, 6,866 of the 10,000 accounts interact with each other via retweets at least once during this time period. There are 6,013 users who retweet others, and 4,957 users who have their tweets retweeted. In Fig.~\ref{fig:retweetsUserDist}, we show the the distribution of retweets among these users.

\subsection{NHS Retweet Network: Aggregating Interactions Versus Using Tie-Decay Networks}
\label{sec:appAggvDecay}

\begin{figure}[htp]
  \centering
  \includegraphics[width=.45\textwidth]{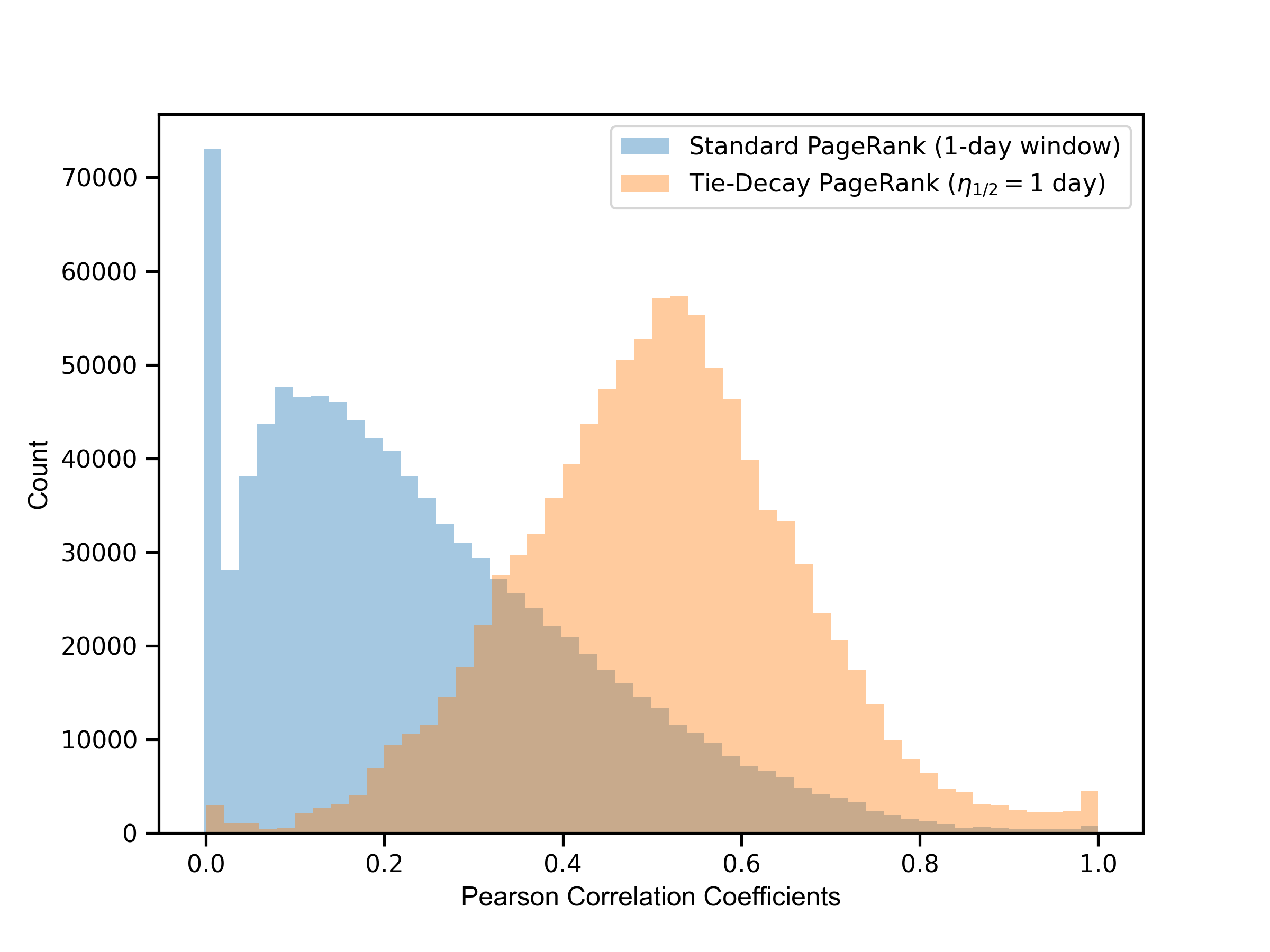}
  \caption{(Color online) 
  Distributions of Pearson correlation coefficients for the correlation matrices in Fig~\ref{fig:NHSCorrMatrices}. We show the distribution for the standard PageRank score with time windows of 1 day in blue and the distribution for tie-decay PageRank with a half-life of $\eta_{1/2} = 1$ day in orange.
  }
  \label{fig:NHSCorrMatDists}
\end{figure}


\pagebreak

\section*{Author Biographies}

\begin{center}
\includegraphics[width=.30\textwidth]{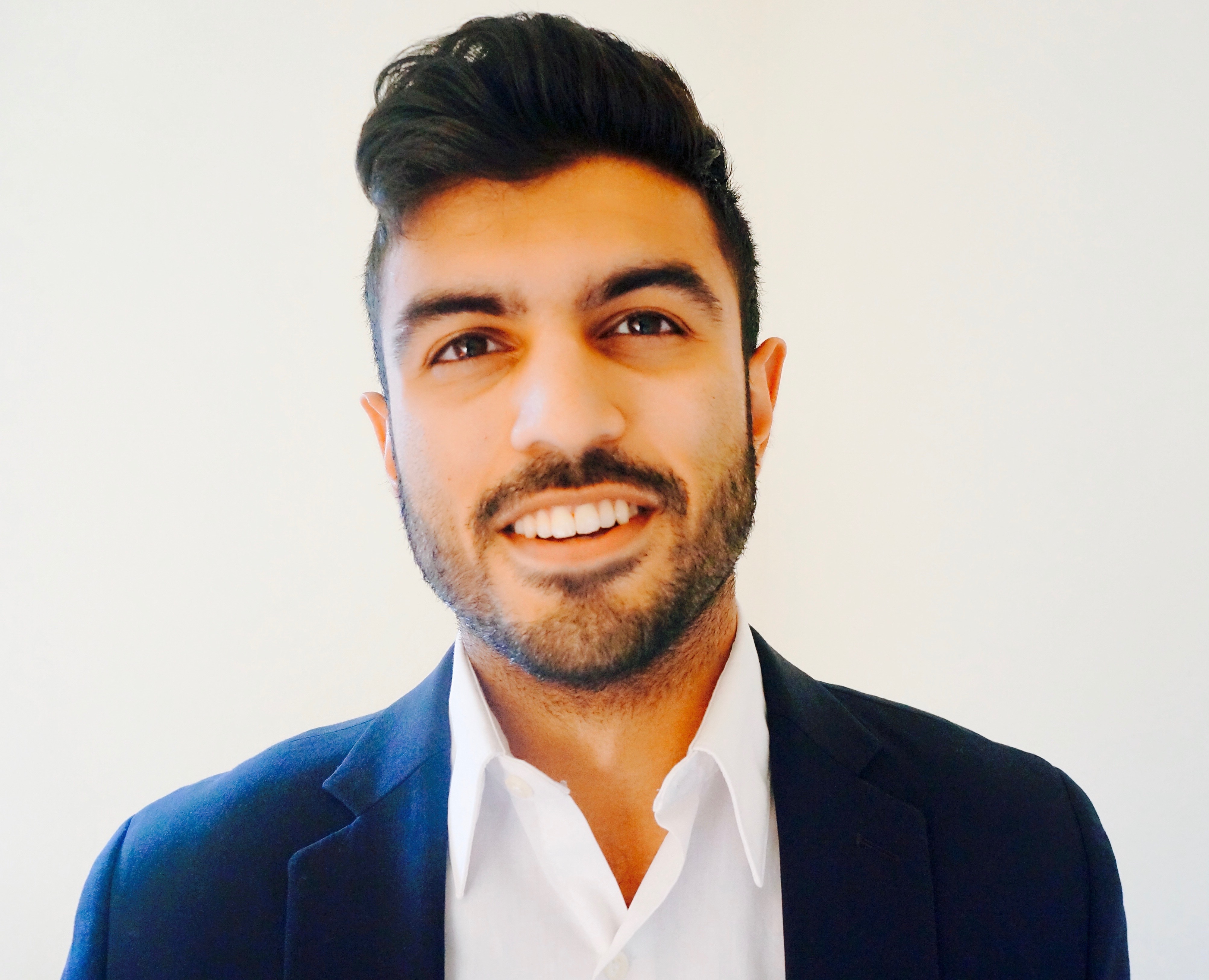}
\end{center}
Walid Ahmad received his B.Sc. in Applied Mathematics from Columbia University
and his M.Sc. in Mathematical Modelling and Scientific Computing from the
University of Oxford. He is currently a machine-learning research engineer
at Reverie Labs, where he works on machine-learning methods for 
computational drug discovery.\\

\begin{center}
\includegraphics[width=.30\textwidth]{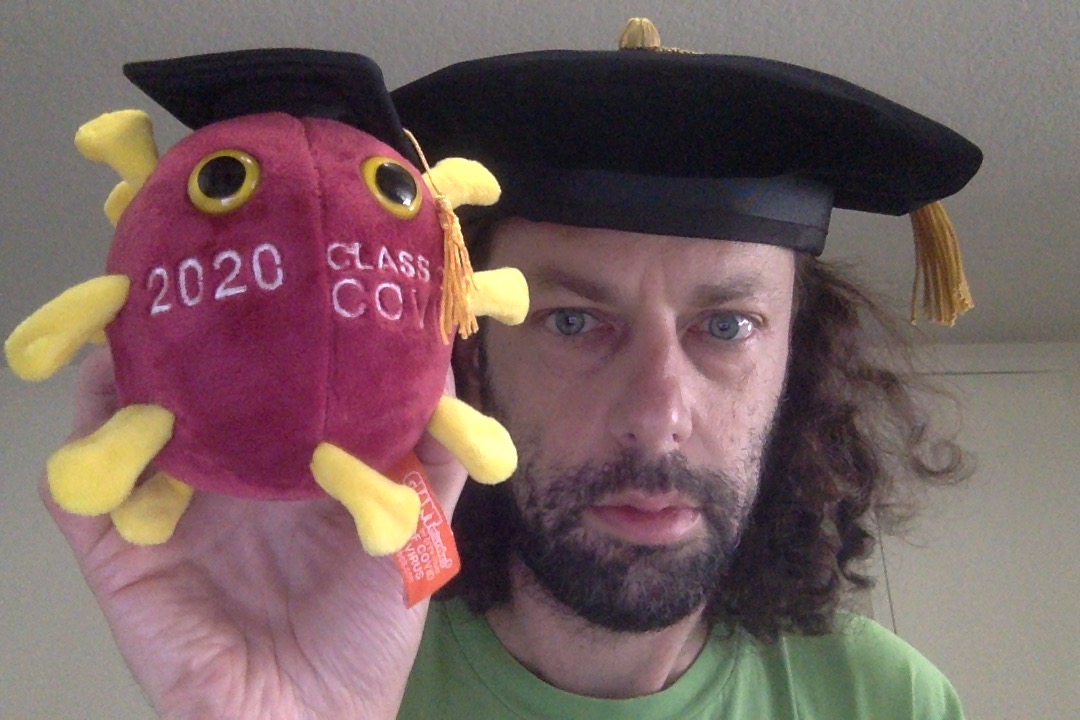}
\end{center}
Mason A. Porter is a professor in the Department of Mathematics at University of
California, Los Angeles. His research interests are in networks,
complex systems, and nonlinear systems. He is a Fellow of the American
Mathematical Society, the American Physical Society, and the Society
for Industrial and Applied Mathematics.\\

\begin{center}
\includegraphics[width=.30\textwidth]{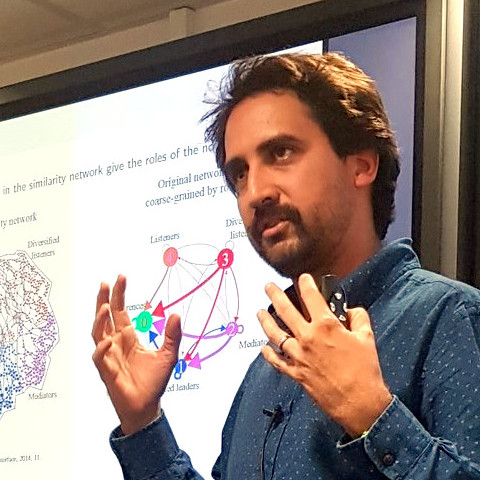}\\
\end{center}
Mariano Beguerisse-D\'iaz is a senior research scientist at Spotify and a visiting
fellow at the University of Oxford. He has held research fellowships
at the University of Oxford and Imperial College London. His research
interests include network science, data science, mathematical biology,
recommendation systems, computational social science, and mathematical
modeling.






\end{document}